# Effective Sample Size for Importance Sampling based on discrepancy measures

L. Martino⋆, V. Elvira⋄, F. Louzada⋆
⋆ Universidade de São Paulo, São Carlos (Brazil).
⋄ Universidad Carlos III de Madrid, Leganés (Spain).

**Abstract**

The Effective Sample Size (ESS) is an important measure of efficiency of Monte Carlo methods such as Markov Chain Monte Carlo (MCMC) and Importance Sampling (IS) techniques. In the IS context, an approximation $\widehat{ESS}$ of the theoretical ESS definition is widely applied, involving the inverse of the sum of the squares of the normalized importance weights. This formula, $\widehat{ESS}$, has become an essential piece within Sequential Monte Carlo (SMC) methods, to assess the convenience of a resampling step. From another perspective, the expression $\widehat{ESS}$ is related to the Euclidean distance between the probability mass described by the normalized weights and the discrete uniform probability mass function (pmf). In this work, we derive other possible ESS functions based on different discrepancy measures between these two pmfs. Several examples are provided involving, for instance, the geometric mean of the weights, the discrete entropy (including the *perplexity* measure, already proposed in literature) and the Gini coefficient among others. We list five theoretical requirements which a generic ESS function should satisfy, allowing us to classify different ESS measures. We also compare the most promising ones by means of numerical simulations.

## 1 Introduction

Sequential Monte Carlo (SMC) methods (a.k.a., particle filtering algorithms) are important tools for Bayesian inference [10], extensively applied in signal processing [9, 17, 29, 26] and statistics [11, 30, 35]. A key point for the success of a SMC method is the use of resampling procedures, applied for avoiding the degeneracy of the importance weights [9, 11]. However, the application of resampling yields loss of diversity in the population of particles and entails an additional computational cost [9, 12], [6, page 21]. Furthermore, resampling limits the parallel implementation of the filter (since it needs the information of all the weights at a specific iteration). Hence, one would desire to perform resampling steps parsimoniously, only when it is strictly required [12, pages 13 and 15]. This adaptive implementation of the resampling procedure needs the use of the concept of *Effective Sample Size* (ESS) of a set of weighted samples [9, 23, 35].

The ESS is a measure of the efficiency of different Monte Carlo methods, such as Markov Chain Monte Carlo (MCMC) and Importance Sampling (IS) techniques [4, 15, 23, 35, 25, 27]. ESS is theoretically defined as the equivalent number of independent samples generated directly form the target distribution, which yields the same efficiency in the estimation obtained by the MCMC or IS algorithms. Thus, a possible mathematical definition [15, 21] considers the ESS as a function proportional to the ratio between the variance of the ideal Monte Carlo estimator (drawing samples directly from the target) over the variance of the estimator obtained by MCMC or IS techniques, using the same number of samples in both estimators.

The most common choice to approximate this theoretical ESS definition in IS is the formula $\widehat{ESS} = \frac{1}{\sum_{n=1}^{N} \bar{w}_n^2}$, which involves (only) the normalized importance weights $\bar{w}_n$, $n = 1, \ldots, N$ [9, 11, 35, 22]. This expression is obtained by several approximations of the initial theoretical definition so that, $\widehat{ESS}$ provides an accurate estimation of the ESS values (given by the theoretical definition) only in specific cases. For this reason other ESS expressions have also been proposed, e.g., the *perplexity*, involving the discrete entropy of the weights [7] has been suggested in [2]; see also [35, Chapter 4], [12, Section 3.5], [18]. The discrete entropy has been also considered in order to design criteria for adaptive resampling schemes in [32, Section 2.3], [31]. More recently, other alternative formulas $\widehat{ESS}$ have also been analyzed in [18]. In [37], a *conditional* $\widehat{ESS}$ formula is introduced in order to study similarities between successive pdfs within a sequence of densities.

However, the ESS approximation $\widehat{ESS} = \frac{1}{\sum_{n=1}^{N} \bar{w}_n^2}$ is widely used in practice, and it generally provides good performance. Furthermore, several theoretical studies related to $\widehat{ESS}$ can be found in literature (e.g., see [1, 36, 32, 31]). It is possible to show that $\widehat{ESS}$ is also related to the discrepancy between the probability mass function (pmf) defined by the normalized weights $\bar{w}_n$, $n = 1, \ldots, N$, and the uniform pmf $\mathcal{U}\{1, 2, \ldots, N\}$. When the pmf defined by $\bar{w}_n$ is close to the uniform pmf $\mathcal{U}\{1, 2, \ldots, N\}$, $\widehat{ESS}$ provides high values otherwise, when the pmf defined by $\bar{w}_n$ is concentrated mainly in one weight, $\widehat{ESS}$ provides small values. More specifically, we show that $\widehat{ESS}$ is related to the Euclidean distance between these two pmfs.

It is possible to obtain other ESS functions based on different discrepancy measures, as we show in this work. We describe and discuss five requirements, three strictly needed and two welcome conditions, that a *Generalized ESS* (G-ESS) function should satisfy. Several examples, involving for instance the geometric mean, discrete entropy [7] and the Gini coefficient [16, 24] of the normalized weights, are presented. Additionally, four families of proper G-ESS functions are designed. We classify the novel G-ESS functions (including also the perplexity measure [2, 35]) according to the conditions fulfilled. We focus on the G-ESS functions which satisfy all the desirable conditions and compare them by means of numerical simulations. This analysis shows that different G-ESS expressions present interesting features from a theoretical and practical point of view and it can be considered valid alternatives of the standard formula $\frac{1}{\sum_{n=1}^{N} \bar{w}_n^2}$.

The rest of the paper is organized as follows. Section 2 recalls the required background material. In Section 3, we highlight that the standard formula $\widehat{ESS} = \frac{1}{\sum_{n=1}^{N} \bar{w}_n^2}$ is related to the Euclidean distance between two pmfs. The definition of a generalized ESS function is given in Section 4, and novel ESS families are introduced in Section 5. Section 6 provides several numerical simulations. Finally, Section 7 contains some brief conclusions.

## 2 Effective Sample Size for Importance Sampling

Let us denote the target pdf as $\bar{\pi}(\mathbf{x}) \propto \pi(\mathbf{x})$ (known up to a normalizing constant) with $\mathbf{x} \in \mathcal{X}$. Moreover, we consider the following integral involving $\bar{\pi}(\mathbf{x})$ and a square-integrable (w.r.t. $\bar{\pi}$) function $h(\mathbf{x})$,

$$I = \int_{\mathcal{X}} h(\mathbf{x})\bar{\pi}(\mathbf{x})d\mathbf{x}, \tag{1}$$

that we desire to approximate using a Monte Carlo approach. If we are able to draw $N$ independent samples $\mathbf{x}_1, \ldots, \mathbf{x}_N$ from $\bar{\pi}(\mathbf{x})$, then the Monte Carlo estimator of $I$ is

$$\widehat{I} = \frac{1}{N}\sum_{n=1}^{N} h(\mathbf{x}_n) \approx I, \tag{2}$$

where $\mathbf{x}_n \sim \bar{\pi}(\mathbf{x})$, with $n = 1, \ldots, N$. However, in general, generating samples directly from the target, $\bar{\pi}(\mathbf{x})$, is impossible. Alternatively, we can draw $N$ samples $\mathbf{x}_1, \ldots, \mathbf{x}_N$ from a simpler proposal pdf $q(\mathbf{x})$,[1] and then assign a weight $w_n = \frac{\pi(\mathbf{x}_n)}{q(\mathbf{x}_n)}$, $n = 1, \ldots, N$, to each sample, according to the importance sampling (IS) approach. Defining the normalized weights,

$$\bar{w}_n = \frac{w_n}{\sum_{i=1}^{N} w_i}, \quad n = 1, \ldots, N, \tag{3}$$

then the IS estimator is

$$\widetilde{I} = \sum_{n=1}^{N} \bar{w}_n h(\mathbf{x}_n) \approx I, \tag{4}$$

with $\mathbf{x}_n \sim q(\mathbf{x})$, $n = 1, \ldots, N$. In general, the estimator $\widetilde{I}$ is less efficient than $\widehat{I}$, since the samples are not directly generate by $\bar{\pi}(\mathbf{x})$. In several applications [9, 11, 17, 26], it is necessary to measure in some way the efficiency that we lose using $\widetilde{I}$ instead of $\widehat{I}$. The idea is to define the Effective Sample Size (ESS) as ratio of the variances of the estimators [21],

$$ESS = N\frac{\text{var}_\pi[\widehat{I}]}{\text{var}_q[\widetilde{I}]}. \tag{5}$$

**Remark 1.** *The ESS value in* (5) *can be interpreted as the number of independent samples drawn directly from the target $\bar{\pi}$ required in order to obtain an estimator $\widehat{I}$ with a variance equal to $var_q[\widetilde{I}]$.*

Namely, ESS represents the number of samples from $\bar{\pi}$ required to obtain a Monte Carlo estimator $\widehat{I}$ with the same efficiency of the IS estimator $\widetilde{I}$ (considering $q$ as proposal). Heuristically speaking, we can assert that ESS measures how many independent identically distributed (i.i.d.) samples, drawn from $\bar{\pi}$, are equivalent to the $N$ weighted samples, draw from $q$ and weighted according to the ratio $\frac{\pi(\mathbf{x})}{q(\mathbf{x})}$ [5, Section 3].

---

[1]We assume that $q(\mathbf{x}) > 0$ for all $\mathbf{x}$ where $\bar{\pi}(\mathbf{x}) \neq 0$, and $q(\mathbf{x})$ has heavier tails than $\bar{\pi}(\mathbf{x})$.

Finding a useful expression of ESS derived analytically from the theoretical definition above is not straightforward. Different derivations [21, 22], [11, Chapter 11], [35, Chapter 4] proceed using several approximations and assumptions for yielding an expression useful from a practical point of view. A well-known ESS approximation, widely used in literature [11, 23, 35], is

$$\begin{aligned}
\widehat{ESS} &= \frac{1}{\sum_{i=1}^{N} \bar{w}_n^2}, \\
&= \frac{\left(\sum_{i=1}^{N} w_n\right)^2}{\sum_{i=1}^{N} w_n^2}, \\
&\triangleq P_N^{(2)}(\bar{\mathbf{w}}),
\end{aligned} \qquad (6)$$

where we have used the normalized weights $\bar{\mathbf{w}} = [\bar{w}_1, \ldots, \bar{w}_N]$ in the first equality, and the unnormalized ones in the second equality. The reason of using the notation $P_N^{(2)}(\bar{\mathbf{w}})$ will appear clear later (the subindex $N$ denotes the number of weights involved, and the reason of the super-index will be clarified in Section 5). An interesting property of the expression (6) is that

$$1 \leq P_N^{(2)}(\bar{\mathbf{w}}) \leq N. \qquad (7)$$

# 3 $P_N^{(2)}$ as a discrepancy measure

Although, in the literature, $P_N^{(2)}(\bar{\mathbf{w}})$ is often considered a suitable approximation of the theoretical ESS definition, the derivation of $P_N^{(2)}$ [21, 34, 35],[5, Section 3] contains several approximations and strong assumptions. As a consequence, $P_N^{(2)}$ differs substantially from the original definition $ESS = N \frac{\text{var}_\pi[\widehat{I}]}{\text{var}_q[\widehat{I}]}$ in many scenarios (e.g., see the numerical results in Section 6.2). In Appendix A, we list the approximations needed in the derivation of $P_N^{(2)}$ and we also discuss its limitations.

Despite of the previous consideration, the expression $P_N^{(2)}(\bar{\mathbf{w}})$ is widely used in the adaptive resampling context [9, 12], [6, page 21] within population Monte Carlo and particle filtering schemes [3, 11, 14, 17, 26].[2] For this reason, several theoretical studies about $P_N^{(2)}$ can also be found in literature [1, 36, 32, 31], showing that $P_N^{(2)}$ has good theoretical properties (e.g., monitoring $P_N^{(2)}$ is enough to prevent the particle system to collapse [1, 31, 32]).

---

[2] In a standard resampling procedure [12, 6], the indices of the particles employed at the next generation are drawn according to a multinomial distribution defined by the normalized weights $\bar{w}_n = \frac{w_n}{\sum_{i=1}^{N} w_i}$, with $n = 1, \ldots, N$. In order to perform resampling steps adaptively, i.e., only in certain specific iterations, the common practice is to estimate the ESS, using typically the approximation $\widehat{ESS} = P_N^{(2)}$. Afterwards, the approximated value $\widehat{ESS}$ is compared with pre-established threshold $\epsilon N$, with $\epsilon \in [0, 1]$ [11, 12, 6]; if $\widehat{ESS} \leq \epsilon N$, then the resampling is applied.

We believe that one of the reasons of the success of $P_N^{(2)}$ in adaptive resampling is due to its connection with the discrepancy between two pmfs: the pmf defined by the weights $\bar{\mathbf{w}} = [\bar{w}_1, \ldots, \bar{w}_N]$ and the discrete uniform pmf defined by $\bar{\mathbf{w}}^* = \left[\frac{1}{N}, \ldots, \frac{1}{N}\right]$. Roughly speaking, if the vector $\bar{\mathbf{w}}$ is reasonably close to $\bar{\mathbf{w}}^*$, then the resampling is considered unnecessary. Otherwise, the resampling is applied. More precisely, we can show that $P_N^{(2)}$ is related to the Euclidean distance $L_2$ between these two pmfs, i.e.,

$$
\begin{aligned}
||\bar{\mathbf{w}} - \bar{\mathbf{w}}^*||_2 &= \sqrt{\sum_{n=1}^{N}\left(\bar{w}_n - \frac{1}{N}\right)^2} \\
&= \sqrt{\left(\sum_{n=1}^{N} \bar{w}_n^2\right) + N\left(\frac{1}{N^2}\right) - \frac{2}{N}\sum_{n=1}^{N} \bar{w}_n} \\
&= \sqrt{\left(\sum_{n=1}^{N} \bar{w}_n^2\right) - \frac{1}{N}} \\
&= \sqrt{\frac{1}{P_N^{(2)}(\bar{\mathbf{w}})} - \frac{1}{N}}.
\end{aligned}
\tag{8}
$$

Hence, maximizing $P_N^{(2)}$ is equivalent to minimizing the Euclidean distance $||\bar{\mathbf{w}} - \bar{\mathbf{w}}^*||_2$. Thus, it appears natural to consider the possibility of using other discrepancy measures between these pmfs, in order to derive alternative ESS functions. In Appendix C, we show other possible ESS expressions induced by non-Euclidean distances. In the following, we define a generic ESS function through the introduction of five conditions (three of them strictly required, and two welcome conditions), and then we provide several examples.

## 4 Generalized ESS functions

In this section, we introduce some properties that a *Generalized ESS (G-ESS) function*, based only on the information of the normalized weights, should satisfy. Here, first of all, note that any possible G-ESS is a function of the vector of normalized weights $\bar{\mathbf{w}} = [\bar{w}_1, \ldots, \bar{w}_N]$,

$$E_N(\bar{\mathbf{w}}) = E_N(\bar{w}_1, \ldots, \bar{w}_N) : \mathcal{S}_N \to [1, N], \tag{9}$$

where $\mathcal{S}_N \subset \mathbb{R}^N$ represents the *unit simplex* in $\mathbb{R}^N$. Namely, the variables $\bar{w}_1, \ldots, \bar{w}_N$ are subjected to the constrain

$$\bar{w}_1 + \bar{w}_2 + \ldots + \bar{w}_N = 1. \tag{10}$$

### 4.1 Conditions for the G-ESS functions

Below we list five conditions that $E_N(\bar{\mathbf{w}})$ should fulfill to be consider a suitable G-ESS function. The first three properties are strictly necessary, whereas the last two are *welcome* conditions, i.e., no strictly required but desirable (see also classification below):

C1. **Symmetry:** $E_N$ must be invariant under any permutation of the weights, i.e.,

$$E_N(\bar{w}_1, \bar{w}_2, \ldots, \bar{w}_N) = E_N(\bar{w}_{j_1}, \bar{w}_{j_2}, \ldots, \bar{w}_{j_N}), \qquad (11)$$

for any possible set of indices $\{j_1, \ldots, j_N\} = \{1, \ldots, N\}$.

C2. **Maximum condition:** A maximum value is $N$ and it is reached at $\bar{\mathbf{w}}^* = \left[\frac{1}{N}, \ldots, \frac{1}{N}\right]$ (see Eq. (51)), i.e.,

$$E_N(\bar{\mathbf{w}}^*) = N \geq E_N(\bar{\mathbf{w}}). \qquad (12)$$

C3. **Minimum condition**: the minimum value is 1 and it is reached (at least) at the vertices $\bar{\mathbf{w}}^{(j)} = [\bar{w}_1 = 0, \ldots, \bar{w}_j = 1, \ldots, \bar{w}_N = 0]$ of the unit simplex in Eq. (52),

$$E_N(\bar{\mathbf{w}}^{(j)}) = 1 \leq E_N(\bar{\mathbf{w}}). \qquad (13)$$

for all $j \in \{1, \ldots, N\}$.

C4. **Unicity of extreme values:** (*welcome condition*) The maximum at $\bar{\mathbf{w}}^*$ is unique and the the minimum value 1 is reached *only* at the vertices $\bar{\mathbf{w}}^{(j)}$, for all $j \in \{1, \ldots, N\}$.

C5. **Stability - Invariance of the rate $\frac{E_N(\bar{\mathbf{w}})}{N}$:** (*welcome condition*) Consider the vector of weights $\bar{\mathbf{w}} = [\bar{w}_1, \ldots, \bar{w}_N] \in \mathbb{R}^N$ and the vector

$$\bar{\mathbf{v}} = [\bar{v}_1, \ldots, \bar{v}_{MN}] \in \mathbb{R}^{MN}, \quad M \geq 1, \qquad (14)$$

obtained repeating and scaling by $\frac{1}{M}$ the entries of $\bar{\mathbf{w}}$, i.e.,

$$\bar{\mathbf{v}} = \frac{1}{M}\underbrace{[\bar{\mathbf{w}}, \bar{\mathbf{w}}, \ldots, \bar{\mathbf{w}}]}_{M-times}. \qquad (15)$$

Note that, clearly, $\sum_{i=1}^{mN} \bar{v}_i = \frac{1}{M}\left[M \sum_{n=1}^{N} \bar{w}_n\right] = 1$. The invariance condition is expressed as

$$\frac{E_N(\bar{\mathbf{w}})}{N} = \frac{E_{MN}(\bar{\mathbf{v}})}{MN}$$
$$E_N(\bar{\mathbf{w}}) = \frac{1}{M} E_{MN}(\bar{\mathbf{v}}), \qquad (16)$$

for all $M \in \mathbb{N}^+$.

The condition C5 is related to the *optimistic approach* described in Appendix B. For clarifying this point, as an example, let us consider the vectors

$$\bar{\mathbf{w}} = [0, 1, 0],$$
$$\bar{\mathbf{v}}' = \left[0, \frac{1}{2}, 0, 0, \frac{1}{2}, 0\right] = \frac{1}{2}[\bar{\mathbf{w}}, \bar{\mathbf{w}}],$$
$$\bar{\mathbf{v}}'' = \left[0, \frac{1}{3}, 0, 0, \frac{1}{3}, 0, 0, \frac{1}{3}, 0\right] = \frac{1}{3}[\bar{\mathbf{w}}, \bar{\mathbf{w}}, \bar{\mathbf{w}}],$$

with $N = 3$. Following the optimistic approach, we should have $E_N(\bar{\mathbf{w}}) = 1$, $E_{2N}(\bar{\mathbf{v}}') = 2$ and $E_{3N}(\bar{\mathbf{v}}'') = 3$, i.e., the rate $E_N/N$ is invariant

$$\frac{E_N(\bar{\mathbf{w}})}{N} = \frac{E_{2N}(\bar{\mathbf{v}}')}{2N} = \frac{E_{3N}(\bar{\mathbf{v}}'')}{3N} = \frac{1}{3}.$$

## 4.2 Classification of G-ESS functions

We divide the possible G-ESS functions in different categories depending on the conditions fulfilled by the corresponding function (see Table 1). Recall that the first three conditions are strictly required. All the G-ESS functions which satisfy at least the first four conditions, i.e., from C1 to C4, are *proper* functions. All the G-ESS functions which satisfy the first three conditions, C1, C2 and C3 but no C4, are considered *degenerate* functions. When a G-ESS function fulfills the last condition is called *stable*. Thus, the G-ESS functions which satisfy all the conditions, i.e., from C1 to C5, are then *proper* and *stable* whereas, if C4 is not satisfied, they are *degenerate* and *stable*. We can also distinguish two type of degeneracy: *type-1* when $E_N(\bar{\mathbf{w}})$ reaches the maximum value $N$ also in some other point $\bar{\mathbf{w}} \neq \bar{\mathbf{w}}^*$, or *type-2* if $E_N(\bar{\mathbf{w}})$ reaches the minimum value 1 also in some point that is not a vertex.

Table 1: Classification of G-ESS depending of the satisfied conditions.

| Class of G-ESS | C1 | C2 | C3 | C4 | C5 |
|---|---|---|---|---|---|
| *Degenerate (D)* | Yes | Yes | Yes | No | No |
| *Proper (P)* | Yes | Yes | Yes | Yes | No |
| *Degenerate and Stable (DS)* | Yes | Yes | Yes | No | Yes |
| *Proper and Stable (PS)* | Yes | Yes | Yes | Yes | Yes |

## 5 G-ESS families and further examples

We can easily design G-ESS functions fulfilling at least the first three conditions, C1, C2, and C3. As examples, considering a parameter $r \geq 0$, we introduce four families of G-ESS functions which have the following analytic forms

$$P_N^{(r)}(\bar{\mathbf{w}}) = \frac{1}{a_r \sum_{n=1}^{N} (\bar{w}_n)^r + b_r}, \quad r \in \mathbb{R}, \quad D_N^{(r)}(\bar{\mathbf{w}}) = \frac{1}{a_r \left[\sum_{n=1}^{N} (\bar{w}_n)^r\right]^{\frac{1}{r}} + b_r}, \quad r \geq 0,$$

$$V_N^{(r)}(\bar{\mathbf{w}}) = a_r \sum_{n=1}^{N} (\bar{w}_n)^r + b_r, \quad r \in \mathbb{R}, \qquad S_N^{(r)}(\bar{\mathbf{w}}) = a_r \left[\sum_{n=1}^{N} (\bar{w}_n)^r\right]^{\frac{1}{r}} + b_r, \quad r \geq 0,$$

where $a_r, b_r$ are constant values depending on the parameter $r$ (and the corresponding family). The values of the coefficients $a_r, b_r$ can be found easily as solutions of linear systems (see Appendix

D), with equations obtained in order to fulfill the conditions C2 and C3. The resulting G-ESS functions are in general *proper*, i.e., satisfying from C1 to C4 (with some degenerate and stable exceptions). The solutions of the corresponding linear systems are given in Table 2. Replacing these solutions within the expressions of the different families, we obtain

$$P_N^{(r)}(\bar{\mathbf{w}}) = \frac{N^{(2-r)} - N}{(1-N)\sum_{n=1}^{N}(\bar{w}_n)^r + N^{(2-r)} - 1}, \qquad (17)$$

$$D_N^{(r)}(\bar{\mathbf{w}}) = \frac{N^{\frac{1}{r}} - N}{(1-N)\left[\sum_{n=1}^{N}(\bar{w}_n)^r\right]^{\frac{1}{r}} + N^{\frac{1}{r}} - 1}, \qquad (18)$$

$$V_N^{(r)}(\bar{\mathbf{w}}) = \frac{N^{r-1}(N-1)}{1 - N^{r-1}}\left[\sum_{n=1}^{N}\bar{w}_n^r\right] + \frac{N^r - 1}{N^{r-1} - 1}, \qquad (19)$$

$$S_N^{(r)}(\bar{\mathbf{w}}) = \frac{N-1}{N^{\frac{1-r}{r}} - 1}\left[\left(\sum_{n=1}^{N}\bar{w}_n^r\right)^{\frac{1}{r}}\right] + 1 - \frac{N-1}{N^{\frac{1-r}{r}} - 1}, \qquad (20)$$

These families contain different G-ESS functions previously introduced, and also other interesting special cases. Table 3 summarizes these particular cases (jointly with the corresponding classification) corresponding to specific values the parameter $r$. Some of them ($D_N^{(0)}$ and $S_N^{(0)}$) involve the *geometric mean* of the normalized weights,

$$\text{GeoM}(\bar{\mathbf{w}}) = \left[\prod_{n=1}^{N}\bar{w}_n\right]^{1/N}, \qquad (21)$$

other ones ($D_N^{(1)} = P_N^{(1)}$ and $S_N^{(1)} = V_N^{(1)}$) involve the *discrete entropy* [7] of the normalized weights,

$$H(\bar{\mathbf{w}}) = -\sum_{n=1}^{N}\bar{w}_n \log_2(\bar{w}_n), \qquad (22)$$

and others use the number of zeros contained in $\bar{\mathbf{w}}$, $N_Z = \#\{\bar{w}_n = 0, \quad \forall n = 1,\ldots,N\}$. The derivations of these special cases are provided in Appendices D.1 and D.2. Note that Table 3 contains a proper and stable G-ESS function

$$S_N^{(1/2)}(\bar{\mathbf{w}}) = \left(\sum_{n=1}^{N}\sqrt{\bar{w}_n}\right)^2, \qquad (23)$$

not introduced so far. Other examples of G-ESS functions, which do not belong to these families, are given below.

**Example 1.** *The following function*

$$Q_N(\bar{\mathbf{w}}) = -N\sum_{i=1}^{N^+}\bar{w}_i^+ + N^+ + N, \qquad (24)$$

Table 2: G-ESS families and their coefficients $a_r$ and $b_r$.

| $P_N^{(r)}(\bar{\mathbf{w}})$ | $D_N^{(r)}(\bar{\mathbf{w}})$ | $V_N^{(r)}(\bar{\mathbf{w}})$ | $S_N^{(r)}(\bar{\mathbf{w}})$ |
|---|---|---|---|
| $\frac{1}{a_r \sum_{n=1}^{N}(\bar{w}_n)^r + b_r}$ | $\frac{1}{a_r \left[\sum_{n=1}^{N}(\bar{w}_n)^r\right]^{\frac{1}{r}} + b_r}$ | $a_r \sum_{n=1}^{N}(\bar{w}_n)^r + b_r$ | $a_r \left[\sum_{n=1}^{N}(\bar{w}_n)^r\right]^{\frac{1}{r}} + b_r$ |
| $a_r = \frac{1-N}{N^{(2-r)}-N}$ | $a_r = \frac{N-1}{N-N^{\frac{1}{r}}}$ | $a_r = \frac{N^{r-1}(N-1)}{1-N^{r-1}}$ | $a_r = \frac{N-1}{N^{\frac{1-r}{r}}-1}$ |
| $b_r = \frac{N^{(2-r)}-1}{N^{(2-r)}-N}$ | $b_r = \frac{1-N^{\frac{1}{r}}}{N-N^{\frac{1}{r}}}$ | $b_r = \frac{N^r-1}{N^{r-1}-1}$ | $b_r = \frac{N^{\frac{1-r}{r}}-N}{N^{\frac{1-r}{r}}-1}$ |

Table 3: Special cases of the families $P_N^{(r)}(\bar{\mathbf{w}})$, $S_N^{(r)}(\bar{\mathbf{w}})$, $D_N^{(r)}(\bar{\mathbf{w}})$ and $V_N^{(r)}(\bar{\mathbf{w}})$.

| **Parameter:** | $\mathbf{r \to 0}$ | $\mathbf{r \to 1}$ | $\mathbf{r = 2}$ | $\mathbf{r \to \infty}$ |
|---|---|---|---|---|
| $P_N^{(r)}(\bar{\mathbf{w}})$ | $\frac{N}{N_Z+1}$ | $\frac{-N\log_2(N)}{-N\log_2(N)+(N-1)H(\bar{\mathbf{w}})}$ | $\frac{1}{\sum_{n=1}^{N}(\bar{w}_n)^2}$ | $\begin{cases} N, & \text{if } \bar{\mathbf{w}} \neq \bar{\mathbf{w}}^{(i)}, \\ 1, & \text{if } \bar{\mathbf{w}} = \bar{\mathbf{w}}^{(i)}. \end{cases}$ |
| | *Degenerate (type-1)* | *Proper* | *Proper-Stable* | *Degenerate (type-1)* |
| **Parameter:** | $\mathbf{r \to 0}$ | $\mathbf{r = \frac{1}{2}}$ | $\mathbf{r \to 1}$ | $\mathbf{r \to \infty}$ |
| $S_N^{(r)}(\bar{\mathbf{w}})$ | $(N^2-N)\text{GeoM}(\bar{\mathbf{w}})+1$ | $\left(\sum_{n=1}^{N}\sqrt{\bar{w}_n}\right)^2$ | $\frac{N-1}{\log_2(N)}H(\bar{\mathbf{w}})+1$ | $N+1-N\max[\bar{w}_1,\ldots,\bar{w}_N]$ |
| | *Degenerate (type-2)* | *Proper-Stable* | *Proper* | *Proper* |
| **Parameter:** | $\mathbf{r \to 0}$ | $\mathbf{r \to 1}$ | | $\mathbf{r \to \infty}$ |
| $D_N^{(r)}(\bar{\mathbf{w}})$ | $\frac{1}{(1-N)\text{GeoM}(\bar{\mathbf{w}})+1}$ | $\frac{-N\log_2(N)}{-N\log_2(N)+(N-1)H(\bar{\mathbf{w}})}$ | | $\frac{1}{\max[\bar{w}_1,\ldots,\bar{w}_N]}$ |
| | *Degenerate (type-2)* | *Proper* | | *Proper-Stable* |
| **Parameter:** | $\mathbf{r \to 0}$ | $\mathbf{r \to 1}$ | | $\mathbf{r \to \infty}$ |
| $V_N^{(r)}(\bar{\mathbf{w}})$ | $N - N_Z$ | $\frac{N-1}{\log_2(N)}H(\bar{\mathbf{w}})+1$ | | $\begin{cases} N & \text{if } \bar{\mathbf{w}} \neq \bar{\mathbf{w}}^{(i)}, \\ 1, & \text{if } \bar{\mathbf{w}} = \bar{\mathbf{w}}^{(i)}. \end{cases}$ |
| | *Degenerate (type-1)-Stable* | *Proper* | | *Degenerate (type-1)* |

with
$$\{\bar{w}_1^+,\ldots,\bar{w}_{N^+}^+\} = \{\text{all } \bar{w}_n: \quad \bar{w}_n \geq 1/N, \quad \forall n = 1,\ldots,N\},$$
and $N^+ = \#\{\bar{w}_1^+,\ldots,\bar{w}_{N^+}^+\}$, is proper and stable. It is related to the $L_1$ distance between $\bar{\mathbf{w}}$ and $\bar{\mathbf{w}}^*$ as shown in Appendix C.

**Example 2.** *The following functions involving the minimum of the normalized weights,*

$$T_{1,N}(\bar{\mathbf{w}}) = \frac{1}{(1-N)\min[\bar{w}_1,\ldots,\bar{w}_N]+1}, \tag{25}$$
$$T_{2,N}(\bar{\mathbf{w}}) = (N^2-N)\min[\bar{w}_1,\ldots,\bar{w}_N]+1, \tag{26}$$

*are degenerate (type-2) G-ESS measures.*

**Example 3.** *The perplexity function introduced in [2] and also contained in a ESS family studied in [18, pages 13 and 22], is defined as*

$$\text{Per}_N(\bar{\mathbf{w}}) = 2^{H(\bar{\mathbf{w}})}, \tag{27}$$

where

$$H(\bar{\mathbf{w}}) = -\sum_{n=1}^{N} \bar{w}_n \log_2(\bar{w}_n), \tag{28}$$

*is the discrete entropy [7] of the pmf $\bar{w}_n$, $n = 1, \ldots, N$. The perplexity is a proper and stable G-ESS function.*

**Example 4.** *Let us consider the Gini coefficient $G(\bar{\mathbf{w}})$ [16, 24], defined as follows. First of all, we define the non-decreasing sequence of normalized weights*

$$\bar{w}_{(1)} \leq \bar{w}_{(2)} \leq \ldots \leq \bar{w}_{(N)}, \tag{29}$$

*obtained sorting in ascending order the entries of the vector $\bar{\mathbf{w}}$. The Gini coefficient is defined as*

$$G(\bar{\mathbf{w}}) = 2\frac{s(\bar{\mathbf{w}})}{N} - \frac{N+1}{N}, \tag{30}$$

where

$$s(\bar{\mathbf{w}}) = \sum_{n=1}^{N} n \bar{w}_{(n)}. \tag{31}$$

*Then, the G-ESS function defined as*

$$\text{Gini}_N(\bar{\mathbf{w}}) = -NG(\bar{\mathbf{w}}) + N, \tag{32}$$

*is proper and stable.*

**Example 5.** *The following G-ESS function (inspired by the $L_1$ distance),*

$$\text{N-plus}_N(\bar{\mathbf{w}}) = N^+ = \#\big\{\bar{w}_n \geq 1/N, \quad \forall n = 1, \ldots, N\big\}. \tag{33}$$

*is also degenerate (type 2) and stable.*

## 5.1 Summary

In the previous sections, we have found different *stable* G-ESS functions, satisfying at least the conditions C1, C2, C3, and C5. They are recalled in Table 4. The following ordering inequalities

$$D_N^{(\infty)}(\bar{\mathbf{w}}) \leq P_N^{(2)}(\bar{\mathbf{w}}) \leq S_N^{(\frac{1}{2})}(\bar{\mathbf{w}}) \leq V_N^{(0)}(\bar{\mathbf{w}}), \quad \forall \bar{\mathbf{w}} \in \mathcal{S}_N,$$

can be also easily proved.

Table 4: Stable G-ESS functions.

| $D_N^{(\infty)}(\bar{\mathbf{w}})$ | $P_N^{(2)}(\bar{\mathbf{w}})$ | $S_N^{(\frac{1}{2})}(\bar{\mathbf{w}})$ | $V_N^{(0)}(\bar{\mathbf{w}})$ |
|---|---|---|---|
| $\frac{1}{\max[\bar{w}_1,\ldots,\bar{w}_N]}$ | $\frac{1}{\sum_{n=1}^{N}\bar{w}_n^2}$ | $\left(\sum_{n=1}^{N}\sqrt{\bar{w}_n}\right)^2$ | $N - N_Z$ |
| *proper* | *proper* | *proper* | *degenerate (type-1)* |

| $Q_N(\bar{\mathbf{w}})$ | N-plus$_N(\bar{\mathbf{w}})$ | Gini$_N(\bar{\mathbf{w}})$ | Per$_N(\bar{\mathbf{w}})$ |
|---|---|---|---|
| $-N\sum_{i=1}^{N^+}\bar{w}_i^+ + N^+ + N$ | $N^+$ | $-NG(\bar{\mathbf{w}}) + N$ | $2^{H(\bar{\mathbf{w}})}$ |
| *proper* | *degenerate (type-2)* | *proper* | *proper* |

## 5.2 Distribution of the ESS values

An additional feature of the G-ESS measures is related to the distribution of the effective sample size values obtained with a specific G-ESS function, when the vector $\bar{\mathbf{w}}$ is considered as a realization of a random variable uniformly distributed in the unit simplex $\mathcal{S}_N$. Namely, let us consider the random variables $\bar{\mathbf{W}} \sim \mathcal{U}(\mathcal{S}_N)$ and $E = E_N(\bar{\mathbf{W}})$ with probability density function (pdf) $p_N(e)$, i.e.,

$$E \sim p_N(e). \tag{34}$$

Clearly, the support of $p_N(e)$ is $[1, N]$. Studying $p_N(e)$, we can define additional properties for discriminating different G-ESS functions. For instance, in general $p_N(e)$ is not a uniform pdf. Some functions $E_N$ concentrate more probability mass closer to the maximum $N$, other functions closer to the minimum 1. This feature varies with $N$, in general. For $N = 2$, it is straightforward to obtain the expression of the pdf $p_2(e)$ for certain G-ESS functions. Indeed, denoting as $I_1(e)$ and $I_2(e)$ the inverse functions corresponding to the monotonic pieces of the generic function $E_2(\bar{w}_1, 1 - \bar{w}_1) = E_2(\bar{w}_1)$, then we obtain

$$p_2(e) = \left|\frac{dI_1}{de}\right| + \left|\frac{dI_2}{de}\right|, \qquad e \in [1, N], \tag{35}$$

using the expression of transformation of a uniform random variable, defined in $[0, 1]$. Thus, we find that $p_2(e) = \frac{2}{e^2}$ for $D_2^{(\infty)}$ and $p_2(e) = \frac{2}{e^2\sqrt{\frac{2}{e}-1}}$ for $P_2^{(2)}$, for instance. Figure 1 depicts the pdfs $p_2(e)$ for $D_2^{(\infty)}$, $T_{2,2}$ in Eq. (26) and $P_2^{(2)}$ in Eq. (6). We can observe that $P_2^{(2)}$ is *more optimistic* than $D_2^{(\infty)}$ *judging* a set of weighted samples and assigning a value of the effective size, since $p_2(e)$ in this case is unbalanced to the right side close to 2. From a practical point of view, the pdf $p_N(e)$ could be used for choosing the threshold values for the adaptive resampling. The *limiting distribution* obtained for $N \to \infty$,

$$p_\infty(e) = \lim_{N \to \infty} p_N(e), \tag{36}$$

is also theoretically interesting, since it can characterize the function $E_N$. However, it is not straightforward to obtain $p_\infty(e)$ analytically. In Section 6, we approximate different limiting pdfs $p_\infty(e)$ of different G-ESS functions via numerical simulation.

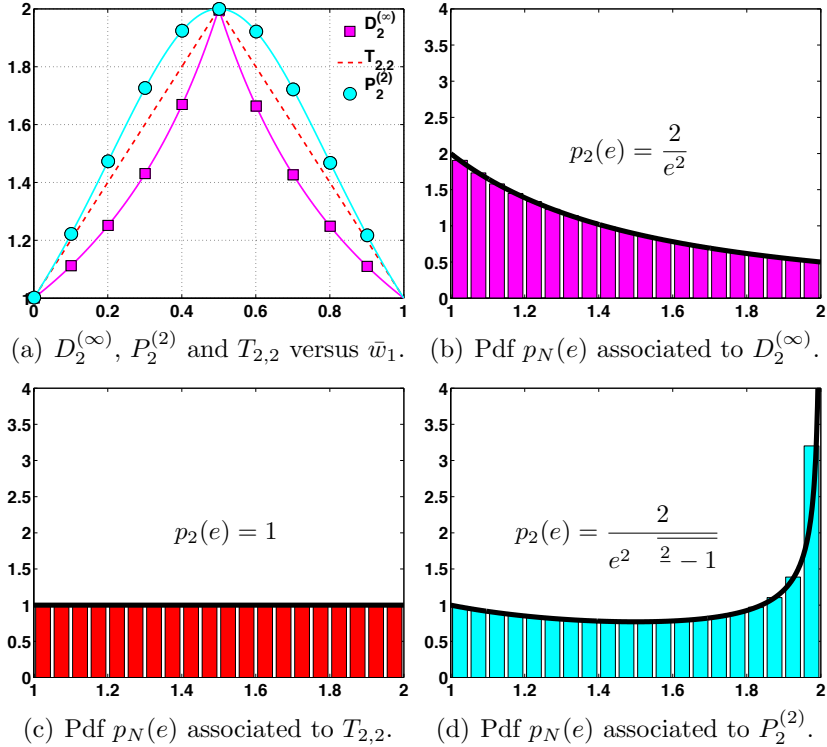

(a) $D_2^{(\infty)}$, $P_2^{(2)}$ and $T_{2,2}$ versus $\bar{w}_1$.  (b) Pdf $p_N(e)$ associated to $D_2^{(\infty)}$.

(c) Pdf $p_N(e)$ associated to $T_{2,2}$.  (d) Pdf $p_N(e)$ associated to $P_2^{(2)}$.

Figure 1: **(a)** G-ESS functions, $D_2^{(\infty)}$ (squares), $P_2^{(2)}$ in Eq. (6) (circles), and $T_{2,2}$ in Eq. (26) (dashed lines), all of them with $N=2$ (then, $\bar{w}_2 = 1-\bar{w}_1$). We can see that $D_2^{(\infty)}$ has a sub-linear increase to the value $N=2$, whereas $P_2^{(2)}$ a super-linear increase. **(b)-(c)-(d)** Pdfs $p_N(e)$ associated to $D_2^{(\infty)}$, $P_2^{(2)}$ and $T_{2,2}$, respectively. For $D_2^{(\infty)}$ (Fig. (b)) more probability mass is located close to 1, whereas for $P_2^{(2)}$ (Fig. (d)), $p_2(e)$ is unbalanced to the right side close to 2.

## 6 Simulations

### 6.1 Analysis of the distribution of ESS values

In this section, we study the distribution $p_N(e)$ of the values of the different G-ESS families. With this purpose, we draw different vectors $\bar{\mathbf{w}}'$ uniformly distributed in the unit simplex $\mathcal{S}_N$, and then we compute the corresponding ESS values (e.g., using the procedure described in [8]). We generate 2000 independent random vectors $\bar{\mathbf{w}}'$ uniformly distributed in the unit simplex $\mathcal{S}_N \subset \mathbb{R}^N$. After that we evaluate the different proper and stable G-ESS functions (summarized in Table 4) at each drawn vector $\bar{\mathbf{w}}'$. The resulting histograms of the rate ESS/N obtained by the different functions are depicted in Figure 2. Figures 2(a)-(c) correspond to $N=50$, whereas (b)-(d) correspond to $N=1000$. Figures 2(a)-(c) show the histograms of the rate corresponding $D_N^{(\infty)}$, $P_N^{(2)}$, $S_N^{(\frac{1}{2})}$, whereas Figures (b)-(d) show the histograms of the rate corresponding $Q_N$, $\text{Gini}_N$ and $\text{Per}_N$. The empirical means and standard deviations for different $N$ are provided in Table 5.

Table 5: Statistics of $\widehat{p}_N(e)$, empirical approximation of $p_N(e)$, corresponding to different G-ESS functions. The greatest standard deviations for a given $N$ are highlighted with boldface.

| Description | N | $D_N^{(\infty)}/N$ | $P_N^{(2)}/N$ | $S_N^{(\frac{1}{2})}/N$ | $Q_N/N$ | $\text{Gini}_N/N$ | $\text{Per}_N/N$ |
|---|---|---|---|---|---|---|---|
| **mean** | 50 | 0.2356 | 0.5194 | 0.7902 | 0.6371 | 0.5117 | 0.6655 |
| | 200 | 0.1776 | 0.5057 | 0.7868 | 0.6326 | 0.5020 | 0.6568 |
| | $10^3$ | 0.1366 | 0.5013 | 0.7858 | 0.6324 | 0.5007 | 0.6558 |
| | $5 \cdot 10^3$ | 0.1121 | 0.5005 | 0.7856 | 0.6322 | 0.5002 | 0.6554 |
| **std** | 50 | 0.0517 | **0.0622** | 0.0324 | 0.0345 | 0.0410 | 0.0492 |
| | 200 | 0.0336 | **0.0341** | 0.0168 | 0.0171 | 0.0204 | 0.0248 |
| | $10^3$ | **0.0213** | 0.0158 | 0.0077 | 0.0077 | 0.0091 | 0.0111 |
| | $5 \cdot 10^3$ | **0.0145** | 0.0071 | 0.0034 | 0.0034 | 0.0040 | 0.0050 |

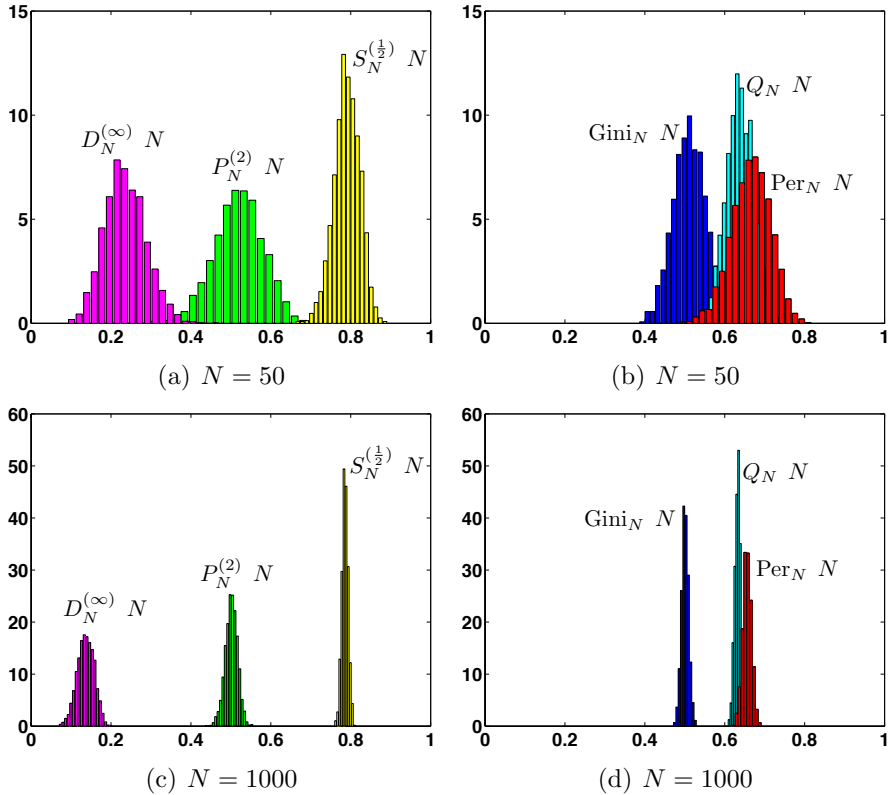

Figure 2: The histograms of the rates $\frac{ESS}{N}$ corresponding to the proper and stable G-ESS functions in Table 4, with $N \in \{50, 1000\}$.

We can observe that all the G-ESS functions concentrate the probability mass of the ESS values around one mode, located in different positions. The variances of these distributions decrease as $N$ grows. The statistical information provided by these histograms can be used for choosing the threshold value in an adaptive resampling scheme. Typically, the condition for applying resampling is

$$E_N(\bar{\mathbf{w}}) \leq \epsilon N,$$

where $0 \leq \epsilon \leq 1$. Namely, the information provided by Table 5 can be useful for choosing $\epsilon$, depending on the used G-ESS function. For instance, Doucet et al. [12, Section 3.5] suggest to use $\epsilon = \frac{1}{2}$ for $P_N^{(2)}$. This suggestion can be explained considering the mean of the ESS values of $P_N^{(2)}$, which is $\approx 0.5$. Moreover, the standard deviation can help us to understand the capability of each formula in differentiating different vectors $\bar{\mathbf{w}}$. The greatest standard deviation for each $N$ is highlighted with boldface. In this sense, $D_N^{(\infty)}$ seems the most "discriminative" for large values of $N$, whereas $P_N^{(2)}$ seems the more convenient for small values of $N$ (however, other studies can suggest the opposite; see below).

## 6.2 Approximation of the theoretical ESS definition

Let us recall the theoretical definition of ESS in Eq. (5),

$$ESS_{var}(h) = N \frac{\text{var}_\pi[\widehat{I}(h)]}{\text{var}_q[\widetilde{I}(h)]}, \qquad (37)$$

where we stress the dependence on the choice of the integrand function $h$. As also discussed in Appendix A, a more convenient definition for small values of $N$ is

$$ESS_{MSE}(h) = N \frac{\text{MSE}_\pi[\widehat{I}(h)]}{\text{MSE}_q[\widetilde{I}(h)]} = N \frac{\text{var}_\pi[\widehat{I}(h)]}{\text{MSE}_q[\widetilde{I}(h)]}. \qquad (38)$$

considering the Mean Square Error (MSE) of the estimators, instead of only the variance. For large values of $N$ the difference between the two definitions is negligible since the bias of $\widetilde{I}$ is virtually zero. In this section, we compute approximately via Monte Carlo the theoretical definitions $ESS_{var}(x)$, $ESS_{MSE}(x)$, and compare with the values obtained with different G-ESS functions. More specifically, we consider a univariate standard Gaussian density as target pdf,

$$\bar{\pi}(x) = \mathcal{N}(x; 0, 1), \qquad (39)$$

and also a Gaussian proposal pdf,

$$q(x) = \mathcal{N}(x; \mu_p, \sigma_p^2), \qquad (40)$$

with mean $\mu_p$ and variance $\sigma_p^2$. Furthermore, we consider different experiment settings:

**S1** In this scenario, we set $\sigma_p = 1$ and vary $\mu_p \in [0, 2]$. Clearly, for $\mu_p = 0$ we have the ideal Monte Carlo case, $q(x) \equiv \bar{\pi}(x)$. As $\mu_p$ increases, the proposal becomes more different from $\bar{\pi}$. We consider the estimation of the expected value of the random variable $X \sim \bar{\pi}(x)$, i.e., we set $h(x) = x$ in the integral of Eq. (1).

**S2** In this case, we set $\mu_p = 1$ and consider $\sigma_p \in [0.23, 4]$. We set $h(x) = x$.

**S3** We fix $\sigma_p = 1$ and $\mu_p \in \{0.3, 0.5, 1, 1.5\}$ and vary the number of samples $N$. We consider again $h(x) = x$.

**S4** In order to analyze the dependence on the choice of $h(x)$ of the theoretical definition (37) and of the numerical results, we consider $h_r(x) = x^r$, $r = 1, \ldots, R = 10$. More specifically, we define the averaged ESS (A-ESS) value,

$$A\text{-}ESS = \frac{1}{R} \sum_{r=1}^{R} ESS_{var}(h_r), \qquad (41)$$

where $ESS_{var}(h_r)$ is given in Eq. (37). First, we set $\sigma_p = 1$, $N \in \{1000, 5000\}$, and vary $\mu_p \in [0, 2]$, as in the setting S1, but we also compute A-ESS in Eq. (41). Then, we set $\sigma_p = 1$, $\mu_p = \{0.3, 1\}$, and vary $N$, similarly as S3.

In the first two cases, we test $N \in \{5, 1000\}$. Figure 3 shows the theoretical ESS curves (approximated via simulations) and the curves corresponding to the proper and stable G-ESS formulas (averaged over $10^5$ independent runs), for the experiment settings S1 and S2. For $N = 1000$, the difference between $ESS_{var}(x)$ and $ESS_{MSE}(x)$ is negligible, so that we only show $ESS_{var}(x)$. For $N = 5$ and S1 we show both curves of $ESS_{var}(x)$ and $ESS_{MSE}(x)$, whereas for $N = 5$ and S2 we only provide $ESS_{MSE}(x)$ since the bias is big for small value of $\sigma_p$ so that it is difficult to obtain reasonable and meaningful values of $ESS_{var}(x)$. Figure 4 and 5 provide the results of the experiment setting S3 and S4, respectively. Note that, for simplicity, in Figure 5 we only show the results of $D_N^{(\infty)}$, $P_N^{(2)}$ and $\text{Gini}_N$, jointly with the theoretical ones, $ESS_{var}(x)$ and A-ESS.

In the setting S1 with $N = 5$ shown Fig. 3(a), first of all we observe that $ESS_{var}(x)$ and $ESS_{MSE}(x)$ are very close when $\mu_p \approx 0$ (i.e., $q(x) \approx \bar{\pi}(x)$) but they differ substantially when the bias increases. In this case, the G-ESS function $\text{Gini}_N$ provides the closest values to $ESS_{var}(x)$, in general. Moreover, $P_N^{(2)}$ and $D_N^{(\infty)}$ also provide good approximations of $ESS_{var}(x)$. Note that $ESS_{var}(x)$ is always contained between $D_N^{(\infty)}$ and $P_N^{(2)}$. In the case S1 with $N = 1000$ shown Fig. 3(b), the formula $P_N^{(2)}$ provides the closest curve to $ESS_{var}(x)$. The G-ESS function $D_N^{(\infty)}$ gives a good approximation when $\mu_p$ increases, i.e., the scenario becomes worse from a Monte Carlo point of view. The G-ESS function $\text{Gini}_N$ provides the best approximation when $\mu_p \in [0, 0.5]$. Again, $ESS_{var}(x)$ is always contained between $D_N^{(\infty)}$ and $P_N^{(2)}$.

In the second scenario S2 with $N = 5$ shown Fig. 3(c), all G-ESS functions are not able to reproduce conveniently the shape of $ESS_{MSE}(x)$. Around to the optimal value of $\sigma_p$, $\text{Gini}_N$ and $P_N^{(2)}$ provide the best approximation of $ESS_{MSE}(x)$. For the rest of value of $\sigma_p$, $D_N^{(\infty)}$ provides the closest results. In the second setting S2 with $N = 1000$ shown Fig. 3(d), $P_N^{(2)}$ seems to emulate better the evolution of $ESS_{var}(x)$. However, $D_N^{(\infty)}$ provides the closest results for small values of $\sigma_p$.

In the experiment setting S3 (see Figure 4), we observe that the behavior of the different G-ESS functions as $N$ grows. When $\mu_p = 0.3$ and $\mu_p = 0.5$, the function $\text{Gini}_N(\bar{\mathbf{w}})$ provides the best approximation of the theoretical definition, i.e., $ESS_{var}(x)$. In particular, with $\mu_p = 0.3$, $\text{Gini}_N(\bar{\mathbf{w}})$ seems to approximate precisely the evolution of $ESS_{var}(x)$. As the proposal differs more to the shape of the target, i.e., for $\mu_p = 1$ and $\mu_p = 1.5$, $D_N^{(\infty)}(\bar{\mathbf{w}})$ becomes the best option. With $\mu_p = 1.5$, $D_N^{(\infty)}(\bar{\mathbf{w}})$ reproduces closely the evolution of $ESS_{var}(x)$. In these last two cases, $\mu_p = 1$ and $\mu_p = 1.5$, $P_N^{(2)}(\bar{\mathbf{w}})$ provides also good performance. We conclude that, in this setup, when the proposal and the target substantially differ, $D_N^{(\infty)}(\bar{\mathbf{w}})$ provides the best results. Roughly speaking, when the shape of proposal is is closer to the shape of target, the function $\text{Gini}_N(\bar{\mathbf{w}})$ provides also good results. Moreover, $\text{Gini}_N(\bar{\mathbf{w}})$ seems to perform better than $P_N^{(2)}(\bar{\mathbf{w}})$ when the number of particles $N$ is small. In intermediate situations, $P_N^{(2)}(\bar{\mathbf{w}})$ seems to be a good compromise. Finally, in the last setting S4, we can observe (see Figure 5) that A-ESS in Eq. (41) is in general smaller than $ESS_{var}(x)$ (which considers only $h(x) = x$). In these experiments, the G-ESS function $D_N^{(\infty)}$ is the closest approximation of A-ESS.

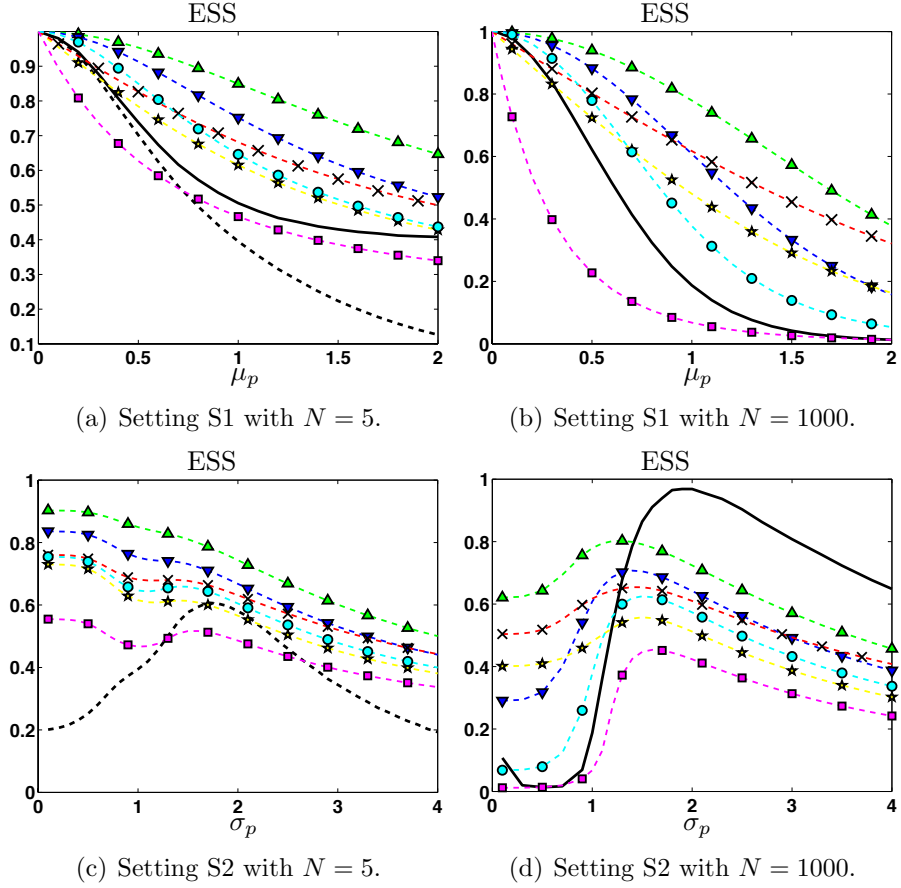

Figure 3: ESS rates corresponding to $ESS_{var}(x)$ (solid line), $ESS_{MSE}(x)$ (dashed line; shown only in (a)-(c)), $P_N^{(2)}$ (circles), $D_N^{(\infty)}$ (squares), $\text{Gini}_N$ (stars), $S_N^{(1/2)}$ (triangles up), $Q_N$ (x-marks), $\text{Per}_N$ (triangles down).

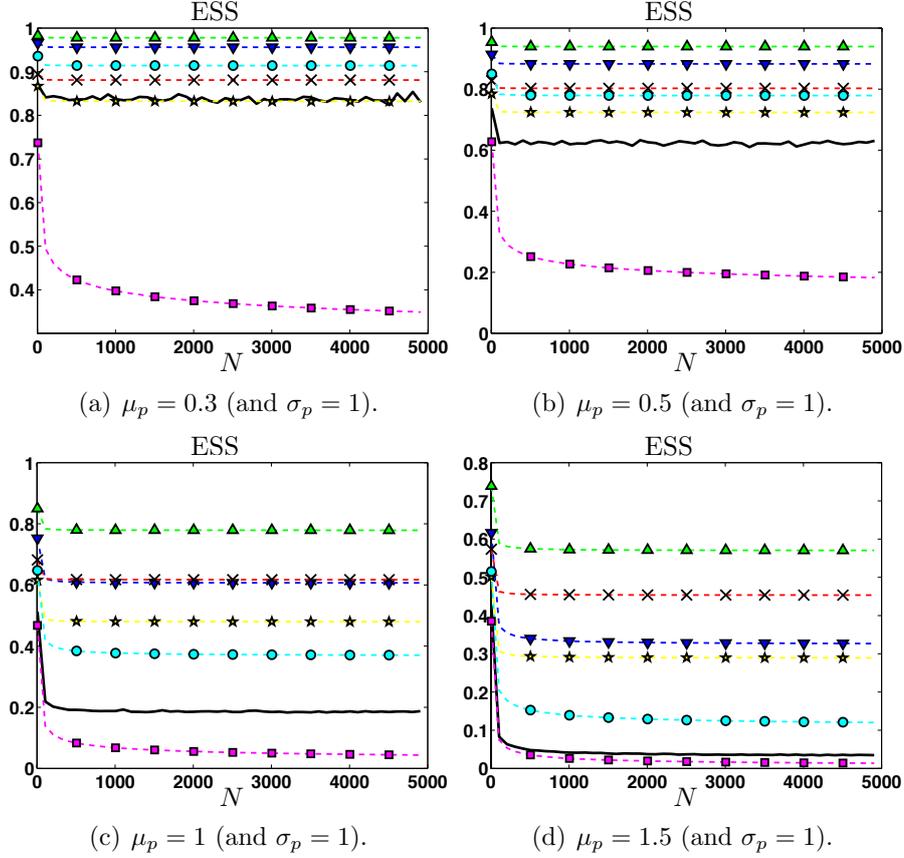

Figure 4: [**Setting S3**] ESS rates as function of $N$, corresponding to the theoretical ESS, i.e., $ESS_{var}(x)$ (solid line), and the G-ESS functions: $P_N^{(2)}$ (circles), $D_N^{(\infty)}$ (squares), $\text{Gini}_N$ (stars), $S_N^{(1/2)}$ (triangles up), $Q_N$ (x-marks), $\text{Per}_N$ (triangles down).

## 6.3 Adaptive Resampling in Particle Filtering

In this example, we apply $P_N^{(2)}$ and $D_N^{(\infty)}$ within a particle filter in order to decide adaptively when performing a resampling step. Specifically, we consider a stochastic volatility model where the hidden state $x_t$ follows an AR(1) process and represents the log-volatility [19] of a financial time series at time $t \in \mathbb{N}$. The equations of the model are given by

$$\begin{cases} x_t = \alpha x_{t-1} + u_t, \\ y_t = \exp\left(\frac{x_t}{2}\right) v_t, \end{cases} \qquad t = 1, \ldots, T. \tag{42}$$

where $\alpha = 0.99$ is the AR parameter, and $u_t$ and $v_t$ denote independent zero-mean Gaussian random variables with variances $\sigma_u^2 = 1$ and $\sigma_v^2 = 0.5$, respectively. Note that $v_t$ is a multiplicative noise. For the sake of simplicity, we implement a standard particle filter (PF) [9, 10, 17] using as propagation equation of the particles exactly the AR(1) process, i.e., the particles $x_{i,t}$'s are propagated as $x_{i,t} \sim p(x_t|x_{i,t-1})$, where $i = 1, \ldots, N$ is the particle index. We set $T = 3000$ and

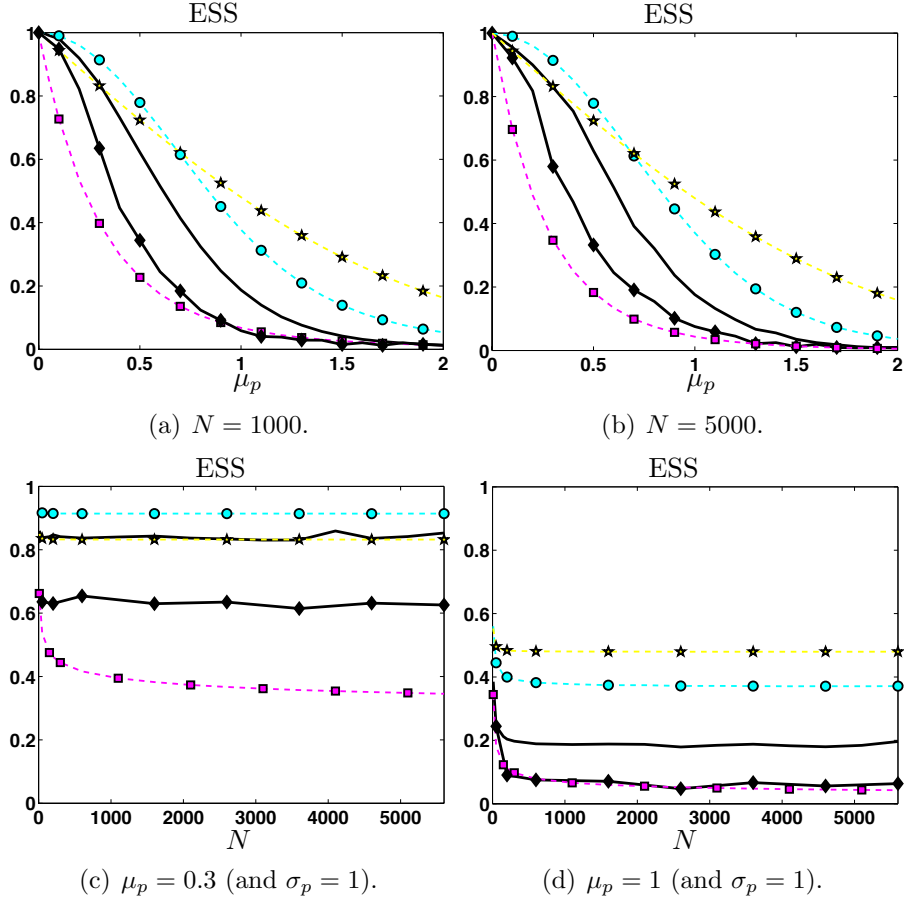

Figure 5: [**Setting S4**] ESS rates corresponding to the theoretical definitions $ESS_{var}(x)$ (solid line), $A\text{-}ESS$ (solid line with rhombuses), and the G-ESS functions $P_N^{(2)}$ (circles), $D_N^{(\infty)}$ (squares), $\text{Gini}_N$ (stars).

$N = 1000$ number of particles. The resampling is performed adaptively, only a certain iterations,

$$\mathcal{T} = \{t_1^*, \ldots, t_r^*\}, \tag{43}$$

where $r = \#\mathcal{T}$ (clearly, $r$ varies in each run). More specifically, denoting as $\bar{\mathbf{w}}_t = [\bar{w}_{1,t}, \ldots, \bar{w}_{N,t}]$ at a specific PF iteration $t$, the conditions for applying the resampling are

$$P_N^{(2)}(\bar{\mathbf{w}}_t) \leq \epsilon_1 N, \qquad D_N^{(\infty)}(\bar{\mathbf{w}}_t) \leq \epsilon_2 N,$$

respectively, where $\epsilon_i \in [0,1]$, $i = 1,2$, are a constant threshold values (with $\epsilon_i = 0$, no resampling is performed; with $\epsilon_i = 1$, the resampling is applied at each iteration).

Let us denote as $\mathcal{T}_1 = \{t_1^*, \ldots, t_{r_1}^*\}$ and $\mathcal{T}_2 = \{\tau_1^*, \ldots, \tau_{r_2}^*\}$ the set of resampling instants obtained by $P_N^{(2)}$ and $D_N^{(\infty)}$, respectively ($r_1 = \#\mathcal{T}_1$ and $r_2 = \#\mathcal{T}_2$). Since $D_N^{(\infty)}(\bar{\mathbf{w}}_t) \geq P_N^{(2)}(\bar{\mathbf{w}}_t)$ for all $\bar{\mathbf{w}}_t \in \mathcal{S}$, and if $\epsilon_1 = \epsilon_2$, using $D_N^{(\infty)}$ we apply more resampling steps than when $P_N^{(2)}$ is used, i.e., $r_2 \geq r_1$ if $\epsilon_1 = \epsilon_2$. However, an equal resampling rate $R$, i.e., the ratio of the averaged number of the performed resampling steps over $T$,

$$R = E\left[\frac{\# \text{ Resampling}}{T}\right] = \frac{1}{T}E[r], \qquad (44)$$

can be obtained using different threshold values $\epsilon_1$ and $\epsilon_2$ for $P_N^{(2)}$ and $D_N^{(\infty)}$. In our case, for obtaining the same resampling rate we need that $\epsilon_1 \geq \epsilon_2$, as shown in Figure 6(a). Note that $0 \leq R \leq 1$.

**Goal.** *Given a resampling rate $R$, our purpose is to discriminate which G-ESS function, between $P_N^{(2)}$ and $D_N^{(\infty)}$, selects the better iteration indices $t^*$'s for applying the resampling steps, i.e., when it is more adequate to apply resampling in order to improve the performance.*

**Results.** We test 100 different values of $\epsilon_1$ and $\epsilon_2$ (we have considered a thin grid of values from 0 to 1 with width 0.01, for both). For each value of $\epsilon_i$, $i = 1, 2$, we run 500 independent simulations of the PF for inferring the sequence $x_{1:t}$, given a sequence of observations $y_{1:T}$ generated according to the model in Eq. (42). Hence, we compute the Mean Square Error (MSE) in the estimation of $x_{1:t}$ obtained by the PF, in each run. Moreover, for each value of $\epsilon_i$, $i = 1, 2$, we calculate the resampling rate $R$ (averaged over the 500 runs). Then, we can plot two curves of averaged MSE versus the resampling rate $R$, corresponding to $P_N^{(2)}$ and $D_N^{(\infty)}$. In this way, we can compare the performance of the PF using the same resampling rate $R$ but obtained with different G-ESS functions, $P_N^{(2)}$ and $D_N^{(\infty)}$. The results are shown in Figure 6(b) in log-log-scale. We can see that, for a given resampling rate $R$, the G-ESS function $D_N^{(\infty)}$ always provides a smaller MSE w.r.t. $P_N^{(2)}$. This confirms that, at least in certain scenarios, $D_N^{(\infty)}$ is a good measure of ESS and it is a valid alternative for $P_N^{(2)}$. Furthermore, the range of useful values of $\epsilon$ in $P_N^{(2)}$ is smaller than in $D_N^{(\infty)}$ as shown in Figure 6(a).

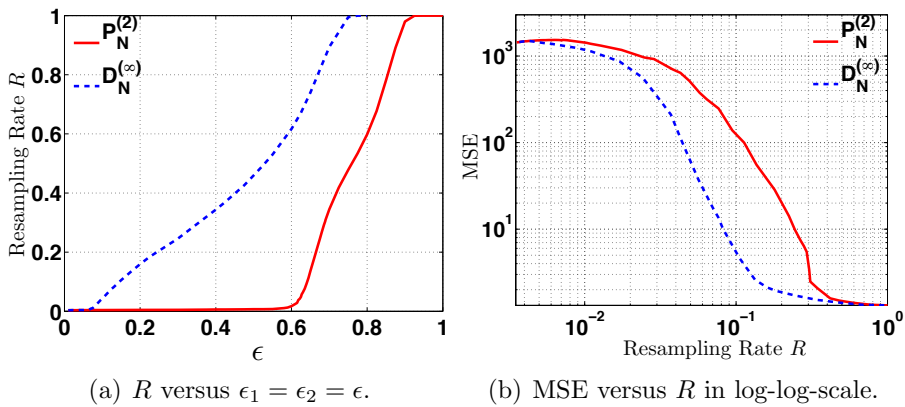

(a) $R$ versus $\epsilon_1 = \epsilon_2 = \epsilon$.  (b) MSE versus $R$ in log-log-scale.

Figure 6: **(a)** Resampling Rate $R$ as function of $\epsilon_1 = \epsilon_2 = \epsilon$ for $P_N^{(2)}$ (solid line) and $D_N^{(\infty)}$ (dashed line). **b** Mean Square Error (MSE) as function of the resampling Rate $R$ for $P_N^{(2)}$ (solid line) and $D_N^{(\infty)}$ (dashed line), in log-log-scale ($N = 1000$ and $T = 3000$).

# 7 Conclusions

In this work, we have proposed new classes of alternative ESS approximations for importance sampling, discussing and testing them from a theoretical and practical point of view. Indeed the novel ESS expressions, jointly with other formulas already presented in literature, have been classified according to five theoretical requirements presented in this work. This classification has allowed to select six different ESS functions which satisfy all these necessary conditions. Then, we have tested them by numerical simulations. Some of them, such as $D_N^{(\infty)}(\bar{\mathbf{w}})$ and $\text{Gini}_N(\bar{\mathbf{w}})$ present interesting features and some benefit, compared to the standard ESS formula $P_N^{(2)}(\bar{\mathbf{w}})$. When the proposal pdf differs substantially to the target density, $D_N^{(\infty)}(\bar{\mathbf{w}})$ provides the best approximations. When the proposal is close to the target, the function $\text{Gini}_N(\bar{\mathbf{w}})$ provides also good results. Moreover, $\text{Gini}_N(\bar{\mathbf{w}})$ seems to perform better than $P_N^{(2)}(\bar{\mathbf{w}})$ when the number of particles $N$ is small. In intermediate scenarios, $P_N^{(2)}(\bar{\mathbf{w}})$ can be also considered a good compromise. Furthermore, $D_N^{(\infty)}(\bar{\mathbf{w}})$ behaves as a "lower bound" for the theoretical ESS definition, as shown in the numerical simulations. The simulation study also provides some useful value for choosing the threshold in an adaptive resampling context. For instance, the results in Table 5 suggest to use of $\epsilon \geq \frac{1}{2}$ for $P_N^{(2)}(\bar{\mathbf{w}})$ (as already noted in [12, Section 3.5]) and $\text{Gini}_N(\bar{\mathbf{w}})$, or $\epsilon \geq 0.11$ for $D_N^{(\infty)}(\bar{\mathbf{w}})$, in the resampling condition $E_N(\mathbf{w}) \leq \epsilon N$. We have also tested $D_N^{(\infty)}$ and $P_N^{(2)}$ within a particle filter for tracking a stochastic volatility variable. The application of G-ESS function $D_N^{(\infty)}$ has provided smaller MSE in estimation w.r.t. $P_N^{(2)}$, considering equal resampling rates (i.e., the number of the performed resampling steps over the total number of iterations of the filter).

# A   Analysis of the theoretical derivation of $P_N^{(2)}$

In the following, we summarize the derivation of $P_N^{(2)}$, that can be partially found in [21] and [23, Section 2.5], stressing the multiple approximations and assumptions:

1. In Section 2, the ESS has been conceptually defined as the ratio between the performance of two estimators, the $\widehat{I}$, where samples are drawn from the target $\bar{\pi}$, and $\widetilde{I}$, the self-normalized IS estimator. The definition $ESS = N\frac{\text{var}_\pi[\widehat{I}]}{\text{var}_q[\widetilde{I}]}$ in Eq. (5) does not take in account the bias of $\widetilde{I}$ (which can be significant for small $N$). Therefore, a more complete definition is

$$ESS = N\frac{\text{MSE}\left[\widehat{I}\right]}{\text{MSE}\left[\widetilde{I}\right]} = N\frac{\text{var}_\pi\left[\widehat{I}\right]}{\text{MSE}\left[\widetilde{I}\right]}, \quad (45)$$

   where we have considered the Mean Square Error (MSE) and we have taken into account that the bias of $\widetilde{I}$. In [21], the derivation starts with the definition in Eq. (5) justifying that *"the bias is of the order of $1/N$ and can be ignored for large $N$"* and that $\widehat{I}$ is unbiased. Indeed, roughly speaking, the squared bias is typically of order $N^{-2}$, which is negligible compared to the variance which is of order $N^{-1}$. Nevertheless, $P_N^{(2)}$ is employed regardless the $N$. Then, the ratio of variances overestimates the theoretical value $ESS = N\frac{\text{MSE}[\widehat{I}]}{\text{MSE}[\widetilde{I}]}$.

2. In the derivation of [21], all the samples are considered to be i.i.d. from a single proposal, i.e. $x_n \sim q(x)$, for $n = 1, ..., N$. Nevertheless, $P_N^{(2)}$ is also used in algorithms which employ multiple proposal pdfs under many different weighting strategies [13].

3. A first delta method is first applied in order to approximate $\text{var}_q[\widetilde{I}]$ in [21, Eq. (6)].

4. The second delta method is applied again to approximate the expectation $\text{E}_\pi[w(\mathbf{x})f(\mathbf{x})^2]$ in [21, Eq. (9)], where $w(\mathbf{x}) = \frac{\pi(\mathbf{x})}{q(\mathbf{x})}$.

5. In the whole derivation, the target is assumed to be normalized (see [21] and [23, Section 2.5]). This is a strong assumption that very rarely occurs in practical scenarios. If this were not the case, the normalizing constant would appear in [21, Eq. (7)-(9)], and therefore also in [21, Eq. (11)-(12)]. As a consequence of the normalized constant assumption, the ESS is approximated as

$$ESS \approx \frac{N}{1 + \text{Var}_q[w(\mathbf{x})]}. \quad (46)$$

   Since $w$ is the unnormalized weights, different scaled version of the target would yield different approximation of the ESS. In order to overcome this problem, it has been proposed (see [35] and the further explanation in [34]) to modify the approximation of Eq. (46) by

$$ESS \approx \frac{N}{1 + \frac{\text{var}_q[w(\mathbf{x})]}{Z^2}} = \frac{N}{1 + \text{CV}^2}, \quad (47)$$

   where CV represents *the coefficient of variation* (also as relative standard deviation) defined as the ratio of the standard deviation $\sqrt{\text{var}_q[w(\mathbf{x})]}$ and the mean $E_q[w(\mathbf{x})] = Z$ [23]. The well-known

$P_N^{(2)}$ can be derived as an empirical approximation of Eq. (47),

$$P_N^{(2)} \approx \frac{N}{1 + \frac{\frac{1}{N}\sum_{n=1}^N w_n^2 - (\frac{1}{N}\sum_{n=1}^N w_n)^2}{(\frac{1}{N}\sum_{n=1}^N w_n)^2}} \qquad (48)$$

$$= \frac{N}{\frac{\frac{1}{N}\sum_{n=1}^N w_n^2}{(\frac{1}{N}\sum_{n=1}^N w_n)^2}} \qquad (49)$$

$$= \frac{1}{\sum_{n=1}^N \bar{w}_n^2}. \qquad (50)$$

Nevertheless, if the target distribution is not assumed to be normalized, the approximation of Eq. (47) is no longer valid. In other words, the metric $P_N^{(2)}$ is approximated with the assumption of $Z = 1$ in the whole derivation, except in the last step where the $Z$ is re-incorporated.

**Consequences**  One consequence of these approximations is that, given the values of $\bar{w}_n$'s, the final formula $P_N^{(2)}$ does not depend directly on the particles $\mathbf{x}_n$, $n = 1, \ldots, N$, which is obviously a drawback since we are trying to measure the effective sample size of the set of weighted particles (see Figure 7).[3] Moreover, $P_N^{(2)}$ is independent from the function $h$, whereas the theoretical definition $ESS = N\frac{\text{var}_\pi[\widehat{I}]}{\text{var}_q[\widetilde{I}]}$ involves $h$. Finally, the inequalities $1 \leq P_N^{(2)} \leq N$ always hold, which can appear an interesting feature after a first examination, but actually it does not encompass completely the theoretical consequences included in the general definition $ESS = N\frac{\text{var}_\pi[\widehat{I}]}{\text{var}_q[\widetilde{I}]}$. Indeed, by this general definition of $ESS$, we have

$$0 \leq ESS \leq B, \qquad B \geq N,$$

i.e., namely $ESS$ can be less than 1, when $\text{var}_q[\widetilde{I}] >> \text{var}_\pi[\widehat{I}]$, and even greater than $N$, when $\text{var}_q[\widetilde{I}] < \text{var}_\pi[\widehat{I}]$: this case occurs when negative correlation is induced among the generated samples [13, 33]. Figure 7 shows the progressive loss of information, first normalizing the weights and then removing the information related to the position of the particles.

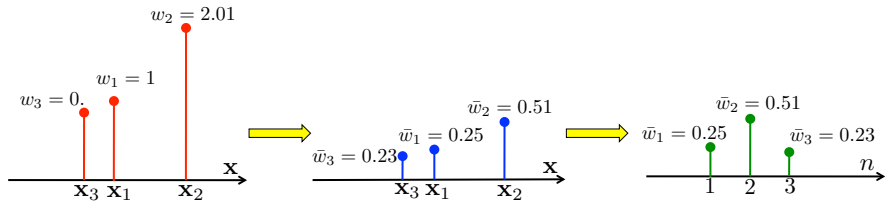

Figure 7: Graphical representation of the loss of statistical information normalizing the weights and ignoring the values of the particles ($N = 3$).

---

[3] As an example, consider the degenerate set of particles where all the samples are the same, i.e., $\mathbf{x}_i = \mathbf{x}_j$ for all $i, j$. In this case, we always have $P_N^{(2)}(\bar{\mathbf{w}}) = N$ which is clearly meaningless (if the target $\pi$ is not a delta function).

# B  The optimism of $P_N^{(2)}(\bar{\mathbf{w}})$

Here, we analyze the behavior of $P_N^{(2)}(\bar{\mathbf{w}})$ in two extreme cases. If all the samples are drawn directly from the target distribution all the weights $w_n$ are equal, so that $\bar{w}_n = \frac{1}{N}$, $n = 1, \ldots, N$. the vector with equal components $\bar{w}_n = \frac{1}{N}$, $n = 1, \ldots, N$, is denoted

$$\bar{\mathbf{w}}^* = \left[\frac{1}{N}, \ldots, \frac{1}{N}\right], \tag{51}$$

Note that the converse is not always true: namely the scenario $\bar{w}_n = \frac{1}{N}$, $n = 1, \ldots, N$, could occur even if the proposal density is different from the target. Hence, in this case, we can assert $ESS \leq N$ (considering independent, non-negative correlated, samples). The other extreme case is

$$\bar{\mathbf{w}}^{(j)} = [\bar{w}_1 = 0, \ldots, \bar{w}_j = 1, \ldots, \bar{w}_N = 0], \tag{52}$$

i.e., $\bar{w}_j = 1$ and $\bar{w}_n = 0$ ( it can occurs only if $\pi(\mathbf{x}_n) = 0$), for $n \neq j$ with $j \in \{1, \ldots, N\}$. The best possible scenario, in this case, is that the $j$-th sample (associate to the weight $\bar{w}_j = 1$) has been generated exactly from $\bar{\pi}(\mathbf{x})$ (hence, with effective sample size equal to 1). Thus, in this case, one can consider $ESS \leq 1$. The function $P_N^{(2)}(\bar{\mathbf{w}})$ employ an optimistic approach for the two extreme cases previously described above:

$$P_N^{(2)}(\bar{\mathbf{w}}^*) = N, \tag{53}$$
$$P_N^{(2)}(\bar{\mathbf{w}}^{(j)}) = 1, \quad \forall j \in \{1, \ldots, N\}. \tag{54}$$

Moreover, considering a vector of type

$$\bar{\mathbf{w}} = \left[0, \frac{1}{C}, \frac{1}{C}, 0, \ldots, 0, 0, \frac{1}{C}, \ldots, 0\right],$$

where only $C$ entries are non-null with the same weight $\frac{1}{C}$, note that

$$P_N^{(2)}(\bar{\mathbf{w}}) = C. \tag{55}$$

Figure 8 summarizes graphically these cases. This approach can appear as a limitation given the previous observations but, using only the information of $\bar{\mathbf{w}}$, appears reasonable.

# C  G-ESS functions induced by non-Euclidean distances

In Section 3, we have pointed out the relationship between $P_N^{(2)}$ and the $L_2$ distance between $\bar{\mathbf{w}}$ and $\bar{\mathbf{w}}^*$. Here, we derive other G-ESS functions given based on non-Euclidean distances.

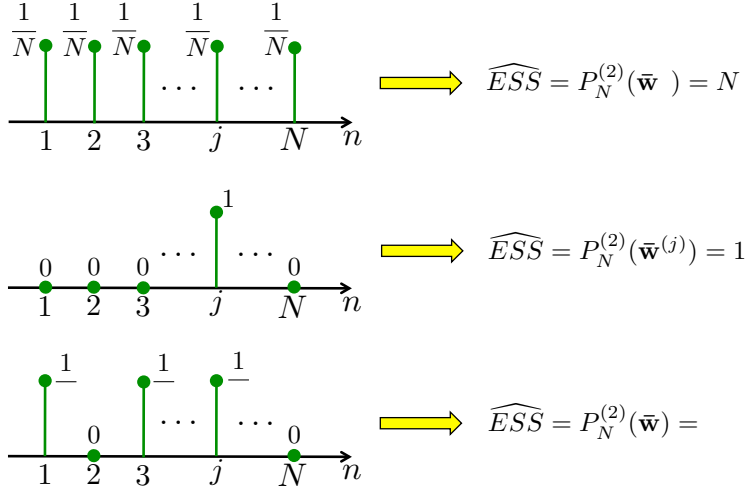

Figure 8: Graphical summary of the optimistic approach employed by $P_N^{(2)}(\bar{\mathbf{w}})$.

**Distance $L_1$** We derive a G-ESS function denoted as $Q_N(\bar{\mathbf{w}})$, induced by the $L_1$ distance. Let us define two disjoint sets of weights

$$\{\bar{w}_1^+, \ldots, \bar{w}_{N^+}^+\} = \{\text{all } \bar{w}_n: \quad \bar{w}_n \geq 1/N, \quad \forall n = 1, \ldots, N\}, \tag{56}$$

$$\{\bar{w}_1^-, \ldots, \bar{w}_{N^-}^-\} = \{\text{all } \bar{w}_n: \quad \bar{w}_n < 1/N, \quad \forall n = 1, \ldots, N\}, \tag{57}$$

where $N^+ = \#\{\bar{w}_1^+, \ldots, \bar{w}_{N^+}^+\}$ and $N^- = \#\{\bar{w}_1^-, \ldots, \bar{w}_{N^+}^-\}$. Clearly, $N^- + N^+ = N$ and $\sum_{i=1}^{N^+} \bar{w}_i^+ + \sum_{i=1}^{N^-} \bar{w}_i^- = 1$. Thus, we can write

$$\begin{aligned}
||\bar{\mathbf{w}} - \bar{\mathbf{w}}^*||_1 &= \sum_{n=1}^{N} \left| \bar{w}_n - \frac{1}{N} \right| \\
&= \sum_{i=1}^{N^+} \left( \bar{w}_i^+ - \frac{1}{N} \right) + \sum_{j=1}^{N^-} \left( \frac{1}{N} - \bar{w}_j^- \right) \\
&= \sum_{i=1}^{N^+} \bar{w}_i^+ - \sum_{i=1}^{N^-} \bar{w}_i^- - \frac{N^+ - N^-}{N}
\end{aligned} \tag{58}$$

and replacing the relationships $\sum_{i=1}^{N^+} \bar{w}_i^+ = 1 - \sum_{i=1}^{N^-} \bar{w}_i^-$ and $N^- = N - N^+$,

$$\begin{aligned}
||\bar{\mathbf{w}} - \bar{\mathbf{w}}^*||_1 &= 2 \left[ \sum_{i=1}^{N^+} \bar{w}_i^+ - \frac{N^+}{N} \right], \\
&= 2 \frac{N \sum_{i=1}^{N^+} \bar{w}_i^+ - N^+}{N}, \\
&= 2 \left[ \frac{N - Q_N(\bar{\mathbf{w}})}{N} \right] + 2,
\end{aligned} \tag{59}$$

where
$$Q_N(\bar{\mathbf{w}}) = -N \sum_{i=1}^{N^+} \bar{w}_i^+ + N^+ + N. \tag{60}$$

Note that $1 \leq Q_N(\bar{\mathbf{w}}) \leq N$, with $Q_N(\bar{\mathbf{w}}^*) = N$ and $Q_N(\bar{\mathbf{w}}^{(i)}) = 1$ for all $i \in \{1, \ldots, N\}$. Maximizing $Q_N$ is equivalent to minimizing the $L_1$ distance between the pmf $\bar{\mathbf{w}}$ and the discrete uniform pmf $\bar{\mathbf{w}}^*$. We remark that this is only one of the possible ESS functions induced by the $L_1$ distance. We choose $Q_N(\bar{\mathbf{w}})$ since it is proper and stable.

**Norm $L_0$**  Interesting G-ESS expressions can be also obtained considering also the distance of the vector $\bar{\mathbf{w}}$ with respect to the null vector containing all zeros as entries (i.e., the norm of $\bar{\mathbf{w}}$). For instance, based on the Hamming distance [7], i.e., we have

$$V_N^{(0)}(\bar{\mathbf{w}}) = N - N_Z, \tag{61}$$

where $N_z$ is the number of zeros in $\bar{\mathbf{w}}$, i.e.,

$$N_Z = \#\{\bar{w}_n = 0, \quad \forall n = 1, \ldots, N\}. \tag{62}$$

Observe that $1 \leq V_N^{(0)}(\bar{\mathbf{w}}) \leq N$ and $V_N^{(0)}(\bar{\mathbf{w}}^*) = N$ and $V_N^{(0)}(\bar{\mathbf{w}}^{(i)}) = 1$ for all $i \in \{1, \ldots, N\}$.

**Norm $L_\infty$**  Other kind of norms can suggest other suitable ESS formulas. For instance,

$$||\bar{\mathbf{w}}||_\infty = \max\left[|\bar{w}_1|, \ldots, |\bar{w}_N|\right] = \frac{1}{D_N(\bar{\mathbf{w}})}, \tag{63}$$

where

$$D_N^{(\infty)}(\bar{\mathbf{w}}) = \frac{1}{\max\left[\bar{w}_1, \ldots, \bar{w}_N\right]}. \tag{64}$$

This G-ESS function has also been recently considered in [18].

# D  Derivation of Generalized ESS families

It is possible to design proper G-ESS fulfilling at least the conditions C1, C2, C3 and C4 (with some degenerate exception), given in the previous section. Below, we show a possible simple procedure but several could be used. Let us consider a function $f(\bar{\mathbf{w}}) : \mathbb{R}^N \to \mathbb{R}$, which satisfies the following properties:

1. $f(\bar{\mathbf{w}})$ is a quasi-concave or a quasi-convex function, with a minimum or a maximum (respectively) at $\bar{\mathbf{w}}^* = \left[\frac{1}{N}, \ldots, \frac{1}{N}\right]$.

2. $f(\bar{\mathbf{w}})$ is symmetric in the sense of Eq. (11).

3. Considering the vertices of the unit simplex $\bar{\mathbf{w}}^{(i)} = \delta(i)$ in Eq. (52), then we also assume

$$f(\bar{\mathbf{w}}^{(i)}) = c,$$

where $c \in \mathbb{R}$ is a constant value, the same for all $i = 1, \ldots, N$.

Let also consider the function $af(\bar{\mathbf{w}}) + b$ obtained as a linear transformation of $f(\bar{\mathbf{w}})$ where $a, b \in \mathbb{R}$ are two constants. Note that, we can always set $a > 0$ if $f(\bar{\mathbf{w}})$ is quasi-concave, or $a < 0$ if $f(\bar{\mathbf{w}})$ is quasi-convex, in order to obtain $g(\bar{\mathbf{w}})$ is always quasi-concave. Hence, we can define the G-ESS function as

$$E_N(\bar{\mathbf{w}}) = \frac{1}{af(\bar{\mathbf{w}}) + b}, \quad \text{or} \quad E_N(\bar{\mathbf{w}}) = af(\bar{\mathbf{w}}) + b, \tag{65}$$

In order to fulfill the properties 2 and 3 in Section 4, recalling $\bar{\mathbf{w}}^* = [\frac{1}{N}, \ldots, \frac{1}{N}]$ and $\bar{\mathbf{w}}^{(i)} = \delta(i)$, we can properly choose the constant values $a, b$ in order to satisfy the following system of $N+1$ equations and two unknowns $a$ and $b$,

$$\begin{cases} af(\bar{\mathbf{w}}^*) + b = \frac{1}{N}, \\ af(\bar{\mathbf{w}}^{(i)}) + b = 1, \quad \forall i \in \{1, \ldots, N\}. \end{cases} \tag{66}$$

or

$$\begin{cases} af(\bar{\mathbf{w}}^*) + b = N, \\ af(\bar{\mathbf{w}}^{(i)}) + b = 1, \quad \forall i \in \{1, \ldots, N\}, \end{cases} \tag{67}$$

respectively. Note that they are both linear with respect to with unknowns $a$ and $b$. Moreover, since $f(\bar{\mathbf{w}}^{(i)}) = c$ for all $i \in \{1, \ldots, N\}$, the system above is reduced to a $2 \times 2$ linear system with solution

$$\begin{cases} a = \frac{N-1}{N[f(\bar{\mathbf{w}}^{(i)}) - f(\bar{\mathbf{w}}^*)]}, \\ b = \frac{f(\bar{\mathbf{w}}^{(i)}) - Nf(\bar{\mathbf{w}}^*)}{N[f(\bar{\mathbf{w}}^{(i)}) - f(\bar{\mathbf{w}}^*)]}. \end{cases} \tag{68}$$

and

$$\begin{cases} a = \frac{N-1}{f(\bar{\mathbf{w}}^*) - f(\bar{\mathbf{w}}^{(i)})}, \\ b = \frac{f(\bar{\mathbf{w}}^*) - Nf(\bar{\mathbf{w}}^{(i)})}{(\bar{\mathbf{w}}^{(i)}) - f(\bar{\mathbf{w}}^{(i)})}. \end{cases} \tag{69}$$

respectively. Below, we derive some special cases of the families $P_N^{(r)}$, $D_N^{(r)}$, $V_N^{(r)}$, and $S_N^{(r)}$ defined in Section 5 and obtained used the procedure above. In these families, we have $f_r(\bar{\mathbf{w}}) = \sum_{n=1}^{N}(\bar{w}_n)^r$ for $P_N^{(r)}$, and $V_N^{(r)}$, and $f_r(\bar{\mathbf{w}}) = \left[\sum_{n=1}^{N}(\bar{w}_n)^r\right]^{1/r}$ for $D_N^{(r)}$ and $S_N^{(r)}$.

## D.1 Special cases of $P_N^{(r)}(\bar{\mathbf{w}})$

In the following, we analyze some special cases of the family $P_N^{(r)}(\bar{\mathbf{w}})$ in Eq. (17):

**Case $r \to 0$.** In this case, the constants in Table 2 reach the values $a_r \to a_0 = -\frac{1}{N}$ and $b_r \to b_0 = \frac{N+1}{N}$. Let us define $0^0 = 0$, considering that $\lim_{r \to 0^+} 0^r = 0$ (i.e., $r$ approaches 0 from the right). With this assumption, Thus, if no zeros are contained in $\bar{\mathbf{w}}$ then $f_0(\bar{\mathbf{w}}) = N$ and $P_N^{(0)}(\bar{\mathbf{w}}) = \frac{1}{Na_0+b_0} = N$, whereas if $\bar{\mathbf{w}}$ contains $N_Z$ zeros, we have $f_0(\bar{\mathbf{w}}) = N - N_Z$ and

$$P_N^{(0)}(\bar{\mathbf{w}}) = \frac{N}{N_Z + 1}, \tag{70}$$

where we recall that $N_Z$ is the number of zero within $\bar{\mathbf{w}}$. Note that, clearly, $P_N^{(0)}(\bar{\mathbf{w}}^{(i)}) = 1$ for all $i \in \{1, \ldots, N\}$, since $N_Z = N - 1$.

**Case $r = 1$.** In this case, $a_r \to \pm\infty$, $b_r \to \mp\infty$, when $r \to 1$. Since $f_r(\bar{\mathbf{w}}) \to 1$ if $r \to 1$, we have an indeterminate form for $g_r(\bar{\mathbf{w}}) = a_r + b_r$ of type $\infty - \infty$. Note that the limit

$$\lim_{r \to 1} P_N^{(r)}(\bar{\mathbf{w}}) = \lim_{r \to 1} \frac{N^{(2-r)} - N}{(1-N)\sum_{n=1}^{N}(\bar{w}_n)^r + N^{(2-r)} - 1},$$

presents an indeterminate form of type $\frac{0}{0}$. Hence, using the L'Hôpital's rule [20], i.e., deriving both numerator and denominator w.r.t. $r$ and computing the limit, we obtain

$$\begin{aligned}
P_N^{(1)}(\bar{\mathbf{w}}) &= \lim_{r \to 1} \frac{-N^{(2-r)}\log(N)}{-N^{(2-r)}\log(N) - (N-1)\sum_{n=1}^{N} \bar{w}_n^r \log(\bar{w}_n)}, \\
&= \frac{-N\log(N)}{-N\log(N) - (N-1)\sum_{n=1}^{N} \bar{w}_n \log(\bar{w}_n)}, \\
&= \frac{-N\frac{\log_2(N)}{\log_2 e}}{-N\frac{\log_2(N)}{\log_2 e} - (N-1)\sum_{n=1}^{N} \bar{w}_n \frac{\log_2(\bar{w}_n)}{\log_2 e}}, \\
&= \frac{-N\log_2(N)}{-N\log_2(N) + (N-1)H(\bar{\mathbf{w}})},
\end{aligned} \tag{71}$$

where we have denoted as $H(\bar{\mathbf{w}}) = -\sum_{n=1}^{N} \bar{w}_n \log_2(\bar{w}_n)$ the discrete entropy of the pmf $\bar{w}_n$, $n = 1, \ldots, N$. Observe that $H(\bar{\mathbf{w}}^*) = \log_2 N$ then $P_N^{(1)}(\bar{\mathbf{w}}) = \frac{-N\log_2(N)}{-\log_2 N} = N$, whereas $H(\bar{\mathbf{w}}^{(i)}) = 0$ (considering $0\log_2 0 = 0$), $P_N^{(1)}(\bar{\mathbf{w}}) = 1$.

**Case $r = 2$.** In this case, $a_2 = 1$ and $b_2 = 0$, hence we obtain

$$P_N^{(2)}(\bar{\mathbf{w}}) = \frac{1}{\sum_{n=1}^{N}(\bar{w}_n)^2}.$$

**Case $r \to \infty$.** We have $a_r \to a_\infty = \frac{N-1}{N}$ and $b_r \to b_\infty = \frac{1}{N}$. If $\bar{\mathbf{w}} \neq \bar{\mathbf{w}}^{(i)}$ for all possible $i \in \{1, \ldots, N\}$, then we have $\lim_{r \to \infty} f_r(\bar{\mathbf{w}}) = 0$ (since $0 < \bar{w}_n < 1$, in this case) and $P_N^{(\infty)}(\bar{\mathbf{w}}) = \frac{1}{b_r} = N$. Otherwise, if $\bar{\mathbf{w}} = \bar{\mathbf{w}}^{(i)}$ for some $i \in \{1, \ldots, N\}$, then $\lim_{r \to \infty} f_r(\bar{\mathbf{w}}) = 1$ (where we have considered $\lim_{r \to \infty} 0^r = 0$ and $\lim_{r \to \infty} 1^r = 1$) and $P_N^{(\infty)}(\bar{\mathbf{w}}) = \frac{1}{a_r + b_r} = 1$. We can summarize both scenarios as

$$P_N^{(\infty)}(\bar{\mathbf{w}}) = \begin{cases} N, & \text{if } \bar{\mathbf{w}} \neq \bar{\mathbf{w}}^{(i)}, \quad \forall i \in \{1, \ldots, N\}, \\ 1, & \text{if } \bar{\mathbf{w}} = \bar{\mathbf{w}}^{(i)}, \quad \forall i \in \{1, \ldots, N\}. \end{cases} \tag{72}$$

## D.2 Special cases of $D_N^{(r)}(\bar{\mathbf{w}})$

Below, we analyze some special cases of the family $D_N^{(r)}(\bar{\mathbf{w}})$ in Eq. (18):

**Case $r \to 0$.** The coefficients of this family given in Table 2 reach the values $a_r \to a_0 = 0$ and $b_r \to b_0 = 1$. In this case, If $\bar{\mathbf{w}} = \bar{\mathbf{w}}^{(i)}$, we have $\lim_{r \to 0} f_r(\bar{\mathbf{w}}) = 1$ (considering again $0^0 = 0$ and $1^\infty = 1$). Whereas, when $\bar{\mathbf{w}}$ is not a vertex, i.e., $\bar{\mathbf{w}} \neq \bar{\mathbf{w}}^{(i)}$, then

$$\lim_{r \to 0} f_r(\bar{\mathbf{w}}) = \lim_{r \to 0} \left[ \sum_{n=1}^N (\bar{w}_n)^r \right]^{\frac{1}{r}} = \infty.$$

so that $\lim_{r \to 0} a_r f_r(\bar{\mathbf{w}})$ has the indeterminate form of type $0 \times \infty$ that can be converted to $\frac{\infty}{\infty}$ as shown below. We can write

$$\lim_{r \to 0} a_r \left[ \sum_{n=1}^N (\bar{w}_n)^r \right]^{\frac{1}{r}} = (1 - N) \lim_{r \to 0} \frac{\left[ \sum_{n=1}^N (\bar{w}_n)^r \right]^{1/r}}{N^{1/r} - N}. \tag{73}$$

Moreover, when $r \to 0$ we have

$$\frac{\left[ \sum_{n=1}^N (\bar{w}_n)^r \right]^{1/r}}{N^{1/r} - N} \approx \frac{\left[ \sum_{n=1}^N (\bar{w}_n)^r \right]^{1/r}}{N^{1/r}} = \left[ \frac{1}{N} \sum_{n=1}^N (\bar{w}_n)^r \right]^{1/r}, \quad \text{when} \quad r \to 0. \tag{74}$$

For $r \to 0$, we can also write

$$(\bar{w}_n)^r = \exp(r \log \bar{w}_n) \approx 1 + r \log \bar{w}_n, \tag{75}$$

where we have used the Taylor expansion of first order of $\exp(r \log \bar{w}_n)$. Replacing $(\bar{w}_n)^r \approx 1 + r \log \bar{w}_n$ inside $\frac{1}{N} \sum_{n=1}^N (\bar{w}_n)^r$ we obtain

$$\frac{1}{N} \sum_{n=1}^N (\bar{w}_n)^r \approx \frac{1}{N} N + r \frac{1}{N} \sum_{n=1}^N \log \bar{w}_n = 1 + r \frac{1}{N} \log \prod_{n=1}^N \bar{w}_n \tag{76}$$

$$= 1 + r \log \left[ \prod_{n=1}^N \bar{w}_n \right]^{\frac{1}{N}}. \tag{77}$$

Thus, we can write

$$\left[\frac{1}{N}\sum_{n=1}^{N}(\bar{w}_n)^r\right]^{1/r} \approx \left[1 + r\log\left[\prod_{n=1}^{N}\bar{w}_n\right]^{\frac{1}{N}}\right]^{1/r}. \tag{78}$$

Moreover, given $x \in \mathbb{R}$, for $r \to 0$ we have also the relationship

$$(1+rx)^{\frac{1}{r}} \to \exp(x),$$

by definition of exponential function. Replacing above, for $r \to 0$,

$$\left[1 + r\log\left[\prod_{n=1}^{N}\bar{w}_n\right]^{\frac{1}{N}}\right]^{1/r} \longrightarrow \exp\left(\log\left[\prod_{n=1}^{N}\bar{w}_n\right]^{\frac{1}{N}}\right) = \left[\prod_{n=1}^{N}\bar{w}_n\right]^{\frac{1}{N}}. \tag{79}$$

Thus, finally we obtain

$$\lim_{r \to 0} a_r \left[\sum_{n=1}^{N}(\bar{w}_n)^r\right]^{\frac{1}{r}} = (1-N)\left[\prod_{n=1}^{N}\bar{w}_n\right]^{1/N}, \tag{80}$$

and

$$\lim_{r \to 0} D_N^{(r)}(\bar{\mathbf{w}}) = \frac{1}{(1-N)\left[\prod_{n=1}^{N}\bar{w}_n\right]^{1/N} + 1}, \tag{81}$$

$$D_N^{(0)}(\bar{\mathbf{w}}) = \frac{1}{(1-N)\text{GeoM}(\bar{\mathbf{w}}) + 1} \tag{82}$$

**Case r= 1.** With a similar procedure used for $P_N^{(1)}$, we obtain $D_N^{(1)}(\bar{\mathbf{w}}) = P_N^{(1)}(\bar{\mathbf{w}})$.

**Case r→ ∞.** In this case, $a_r \to a_\infty = 1$ and $b_r \to b_\infty = 0$ and, since the distance $\left[\sum_{n=1}^{N}(\bar{w}_n)^r\right]^{\frac{1}{r}}$ converges to the $L_\infty$ distance, $\max[\bar{w}_1, \ldots, \bar{w}_N]$, when $r \to \infty$ [28], we obtain $D_N^{(\infty)}(\bar{\mathbf{w}}) = \frac{1}{\max[\bar{w}_1, \ldots, \bar{w}_N]}$.

## D.3 Special cases of $V_N^{(r)}(\bar{\mathbf{w}})$

In the following, we study some special cases of the family $V_N^{(r)}(\bar{\mathbf{w}})$ in Eq. (19):

**Case r→ 0.** The coefficients of this family given in Table 2 are $a_0 = 1$ and $b_0 = 0$. If $\bar{\mathbf{w}}$ does not contain zeros then $f_r(\bar{\mathbf{w}}) = \sum_{n=1}^{N}(\bar{w}_n)^r = N$. Otherwise, assuming $0^0 = 0$, If $\bar{\mathbf{w}}$ contains $N_Z$ zeros, we have $f_r(\bar{\mathbf{w}}) = \sum_{n=1}^{N}(\bar{w}_n)^r = N - N_Z$. Thus, in general, we have

$$V^{(0)}(\bar{\mathbf{w}}) = N - N_Z.$$

**Case r→ 1.** In this, the coefficients $a_r$ and $b_r$ diverge. We can consider the limit

$$\lim_{r \to 1} \left( \frac{N^{r-1}(N-1)}{1 - N^{r-1}} \left[ \sum_{n=1}^{N} \bar{w}_n^r \right] + \frac{N^r - 1}{N^{r-1} - 1} \right) =$$

$$= (N-1) \lim_{r \to 1} \frac{\left[ \sum_{n=1}^{N} \bar{w}_n^r \right] - 1}{1 - N^{r-1}},$$

where we have a indetermination of type $\frac{0}{0}$. Using the L'Hôpital's rule [20], i.e., deriving both numerator and denominator w.r.t. $r$ and computing the limit, we obtain (since $\log x = \frac{\log_2 x}{\log_2 e}$)

$$\lim_{r \to 1} \frac{\left[ \sum_{n=1}^{N} \bar{w}_n^r \right] - 1}{1 - N^{r-1}} = \lim_{r \to 1} \frac{\left[ \sum_{n=1}^{N} (\bar{w}_n^r \log \bar{w}_n) \right]}{-N^{r-1} \log N},$$

$$= \frac{\left[ -\sum_{n=1}^{N} (\bar{w}_n \log \bar{w}_n) \right]}{\log N} = \frac{\left[ -\sum_{n=1}^{N} (\bar{w}_n \log_2 \bar{w}_n) \right]}{(\log_2 e) \frac{\log_2 N}{\log_2 e}},$$

$$= \frac{H(\bar{\mathbf{w}})}{\log_2 N},$$

hence, finally,

$$V_N^{(1)}(\bar{\mathbf{w}}) = (N-1) \frac{H(\bar{\mathbf{w}})}{\log_2 N} + 1. \tag{83}$$

**Case r→ ∞.** The coefficients converge to the values $a_r \to a_\infty = 1 - N$ and $b_r \to b_\infty = N$. If $\bar{\mathbf{w}} \neq \bar{\mathbf{w}}^{(i)}$ then $f_r(\bar{\mathbf{w}}) = \sum_{n=1}^{N} (\bar{w}_n)^r = 0$, so that $V_N^{(\infty)}(\bar{\mathbf{w}}) = b_\infty = N$. Otherwise, If $\bar{\mathbf{w}} = \bar{\mathbf{w}}^{(i)}$, since $0^\infty = 0$ and considering $1^\infty = 1$, then $f_r(\bar{\mathbf{w}}) = \sum_{n=1}^{N} (\bar{w}_n)^r = 1$, so that $V_N^{(\infty)}(\bar{\mathbf{w}}) = a_\infty + b_\infty = 1$.

## D.4 Special cases of $S_N^{(r)}(\bar{\mathbf{w}})$

Let us consider the family $S_N^{(r)}(\bar{\mathbf{w}})$. Four interesting special cases are studied below:

**Case r→ 0.** The coefficients given in Table 2 in this case are $a_r \to a_0 = 0$ and $b_r \to b_0 = 1$. If $\bar{\mathbf{w}} \neq \bar{\mathbf{w}}^{(i)}$ then $f_r(\bar{\mathbf{w}}) \to \infty$, Otherwise, If $\bar{\mathbf{w}} = \bar{\mathbf{w}}^{(i)}$, since $0^\infty = 0$ and considering $1^\infty = 1$, then $f_r(\bar{\mathbf{w}}) \to 1$. With a procedure similar to $D_N^{(0)}$, it is possible to show that

$$S^{(0)}(\bar{\mathbf{w}}) = (N^2 - N)\text{GeoM}(\bar{\mathbf{w}}) + 1. \tag{84}$$

**Case r→ $\frac{1}{2}$.** We have $a_{1/2} = 1$ and $b_{1/2} = 0$. Then, in this case, $S_N^{(\frac{1}{2})}(\bar{\mathbf{w}}) = f_{1/2}(\bar{\mathbf{w}}) = \left( \sum_{n=1}^{N} \sqrt{\bar{w}_n} \right)^2$.

**Case r→ 1.** With a similar procedure used for $V_N^{(1)}$, it is possible to obtain

$$S_N^{(1)}(\bar{\mathbf{w}}) = (N-1)\frac{H(\bar{\mathbf{w}})}{\log_2 N} + 1. \tag{85}$$

**Case r→ ∞.** In this case, $a_r \to a_\infty = -N$, $b_r \to b_\infty = N+1$. Moreover, $f_r(\bar{\mathbf{w}}) \to \max[\bar{w}_1, \ldots, \bar{w}_n]$ [28], so that

$$S_N^{(\infty)}(\bar{\mathbf{w}}) = (N+1) - N\max[\bar{w}_1, \ldots, \bar{w}_n]. \tag{86}$$

Table 3 summarizes all the special cases.